\documentclass[acmsmall,screen]{acmart}
\usepackage{pbox}
\usepackage{prftree}
\usepackage{ amssymb }

\usepackage{microtype}
\usepackage{xspace}

\usepackage{tikz}
\usetikzlibrary{matrix}

\usepackage{colortbl}
\usepackage{listings}
\usepackage{csvsimple}
\usepackage{mfirstuc}
\usepackage{siunitx,array,booktabs}
\usepackage{changepage}
\usepackage{dblfloatfix}    
\usepackage{wrapfig} 
\usepackage{tikzpagenodes}
\usepackage{caption}

\newtheoremstyle{break}
  {}
  {}
  {\itshape}
  {}
  {\bfseries}
  {.}
  {\newline}
  {}

\theoremstyle{break}

\newtheorem{lemma1}{Lemma}[subsection]
\newtheorem{theorem1}[lemma1]{Theorem}
\newtheorem{definition1}[lemma1]{Definition}
\newtheorem{sketchdefinition1}[lemma1]{Definition (sketch)}

\let\qed\relax

\newcommand\void[1]{}

\newdimen\proofadjustabove
\newdimen\proofadjustbelow
\newenvironment{proof}{\setcounter{proofcase}{0}\vspace{\proofadjustabove}\rm{\bf Proof}��\vspace{\proofadjustbelow}\begin{list}{}{\leftmargin=1.5em\labelsep=1.5em\labelwidth=0pt}\item}{\end{list}}
\newenvironment{proofsketch}{\setcounter{proofcase}{0}\vspace{\proofadjustabove}\rm{\bf Proof Sketch}\vspace{\proofadjustbelow}\begin{list}{}{\leftmargin=1.5em\labelsep=1.5em\labelwidth=0pt}\item}{\end{list}}
\newcounter{proofcase}

\newcounter{proofsubcase}

\newcommand{\qed}{\quad\ensuremath{\blacksquare}\parfillskip 0pt\linebreak}
\renewenvironment{proof}{\noindent{\bf Proof:}\hspace*{0.5em}}{\hspace*{\fill}\qed}

\newcommand\x[1]{\ensuremath{\mathit{#1}}\xspace}

\newcommand\K[1]{\ensuremath{\textsf{\sf#1}}}
\newcommand\KK[1]{\ensuremath{\K{\textbf{#1}}}}

\newcommand\Kto{\ensuremath{\rightarrow}\xspace}

\DeclareTextCommand{\_}{OT1}{\leavevmode \kern.06em\vbox{\hrule width.3em}}

\definecolor{hilite}{rgb}{0.7,0,0.5}
\newcommand{\hilite}[1]{\color{purple}#1\color{black}}

\definecolor{shade}{rgb}{0.6,0.6,0.6}

\definecolor{hishade}{rgb}{1,0.4,0.4}

\newcommand{\wasm}{\textsf{Wasm}\xspace}
\newcommand{\simcfg}{\sim_{\textnormal{\textbf{c}}}}
\newcommand{\simact}{\sim_{\textnormal{\textbf{a}}}}
\newcommand{\simstr}{\sim_{\textnormal{\textbf{s}}}}
\newcommand{\simval}{\sim_{\textnormal{\textbf{v}}}}
\newcommand{\simexp}{\sim_{\textnormal{\textbf{e}}}}
\newcommand{\succtr}{\succ_{\textnormal{\textbf{tr}}}}
\newcommand{\ctwasm}{\textsf{CT-Wasm}\xspace}
\newcommand{\strip}{\textsf{ct2wasm}\xspace}
\newcommand{\inferer}{\textsf{wasm2ct}\xspace}
\newcommand{\js}{JavaScript\xspace}

\newcommand{\para}[1]{\smallskip\noindent\textbf{{#1.}\xspace}}

\newcommand{\dudect}{\textsf{dudect}\xspace}
\newcommand{\ctverif}{\textsf{ct-verif}\xspace}

 \renewcommand\_{{\textunderscore}}

\setcopyright{rightsretained}
\acmPrice{}
\acmDOI{10.1145/3290390}
\acmYear{2019}
\copyrightyear{2019}
\acmJournal{PACMPL}
\acmVolume{3}
\acmNumber{POPL}
\acmArticle{77}
\acmMonth{1}

\usepackage{booktabs}   
\usepackage{subcaption} 

\begin{document}






\title[CT-Wasm: Type-Driven Secure Cryptography for the Web Ecosystem]{CT-Wasm: Type-Driven Secure Cryptography \\for the Web Ecosystem}         


\author{Conrad Watt}
\orcid{0000-0002-0596-877X}             
\affiliation{
  \institution{University of Cambridge}            
  \country{UK}                    
}
\email{conrad.watt@cl.cam.ac.uk}          

\author{John Renner}
\affiliation{
  \institution{University of California San Diego}            
  \country{USA}                    
}
\email{jmrenner@eng.ucsd.edu}          

\author{Natalie Popescu}
\affiliation{
  \institution{University of California San Diego}            
  \country{USA}                    
}
\email{npopescu@ucsd.edu}          

\author{Sunjay Cauligi}
\affiliation{
  \institution{University of California San Diego}            
  \country{USA}                    
}
\email{scauligi@eng.ucsd.edu}          

\author{Deian Stefan}
\affiliation{
  \institution{University of California San Diego}            
  \country{USA}                    
}
\email{deian@cs.ucsd.edu}          

\begin{abstract}
A significant amount of both client and server-side cryptography is implemented
in JavaScript.
Despite widespread concerns about its security, no other language
has been able to match the convenience that comes from its ubiquitous support
on the ``web ecosystem''---the wide variety of technologies that collectively underpins
the modern World Wide Web.
With the introduction of the new WebAssembly bytecode language (Wasm) into the web ecosystem, we
have a unique opportunity to advance a principled alternative to existing JavaScript
cryptography use cases which does not compromise this convenience.

We present Constant-Time WebAssembly (CT-Wasm), a type-driven, strict extension
to WebAssembly which facilitates the verifiably secure implementation of
cryptographic algorithms. 
CT-Wasm's type system ensures that code written in CT-Wasm is both information
flow secure and resistant to timing side channel attacks; like base Wasm, these
guarantees are verifiable in linear time.
Building on an existing Wasm mechanization, we mechanize the full CT-Wasm
specification, prove soundness of the extended type system, implement a
verified type checker, and give several proofs of the language's security
properties.

We provide two implementations of CT-Wasm: an OCaml reference interpreter and a
native implementation for Node.js and Chromium that extends Google's V8 engine.
We also implement a CT-Wasm to Wasm rewrite tool that allows developers to reap
the benefits of CT-Wasm's type system today, while developing cryptographic
algorithms for base Wasm environments.
We evaluate the language, our implementations, and supporting tools by porting
several cryptographic primitives---Salsa20, SHA-256, and TEA---and the full
TweetNaCl library.
We find that CT-Wasm is fast, expressive, and generates code that we
experimentally measure to be constant-time.
\end{abstract}

\ccsdesc[500]{Security and privacy~Cryptography}
\ccsdesc[300]{Security and privacy~Formal methods and theory of security}
\ccsdesc[300]{Security and privacy~Web application security}
\ccsdesc[300]{Theory of computation~Operational semantics}
\ccsdesc[300]{Software and its engineering~Assembly languages}

\keywords{WebAssembly, cryptography, constant-time, information flow control}

\maketitle

\fancyfoot{}
\thispagestyle{empty}

\section{Introduction}
\label{sec:intro}
 
When implementing a cryptographic algorithm, functional correctness alone is
not sufficient.
It is also important to ensure properties about information flow that take into
account the existence of \emph{side channels}---ways in which information can
be leaked as side-effects of the computation process.
For example, the duration of the computation itself can be a side channel,
since an attacker could compare different executions to infer which program
paths were exercised, and work backwards to determine information about secret
keys and messages.

Writing code that does not leak information via side channels is daunting even with
complete control over the execution environment, but in recent years an even more
challenging environment has emerged---that of \textit{in-browser cryptography}---the
implementation of cryptographic algorithms in a user's browser using \js.
Modern \js runtimes are extremely complex software systems,
incorporating just-in-time (JIT) compilation and garbage collection (GC) techniques that almost
inherently expose timing side-channels~\cite{jitsec, Oren:2015:SSP:2810103.2813708,
VanGoethem:2015:CST:2810103.2813632}.
Even worse, much of the \js cryptography used in the wild is implemented by
``unskilled cryptographers''~\cite{unskilled} who do not account for even the most basic timing
side channels.
It is dangerous enough that unsecure, in-browser cryptography has become commonplace
on the web, but the overwhelming popularity of JavaScript as a development
language across all platforms~\cite{ibmwhitepaper} has driven adoption of
JavaScript cryptography on the server-side as well. With multiple JavaScript crypto libraries served by the NPM package
manager alone having multiple-millions of weekly downloads~\cite{npm-crypto-browserify, npm-tweetnacl, npm-elliptic,
npm-pbkdf2}), many of the issues noted above
are also exposed server-side.

To fundamentally address the state of crypto in the web ecosystem, a solution
must simultaneously compete with the apparent convenience of
JavaScript crypto for developers while having better security characteristics.
Modifying complex, ever-evolving \js engines to protect \js code from leakage via timing channels would
be a labyrinthine task.
Luckily, this is not necessary: all major browsers recently added support for
WebAssembly (\wasm)~\cite{wasm_landing,Haas:2017:BWU:3062341.3062363}.

\wasm is a low-level bytecode language.
This alone provides a firmer foundation for cryptography than \js: \wasm's
close-to-the-metal instructions give us more confidence in its timing
characteristics than \js's unpredictable optimizations.
WebAssembly also distinguishes itself through its strong, static type
system, and principled design.
Specifically, \wasm has a formal small-step
semantics~\cite{Haas:2017:BWU:3062341.3062363};
well-typed \wasm programs enjoy standard \textit{progress} and
\textit{preservation} properties~\cite{Wright:1994:SAT:191905.191909}, which
have even been mechanically verified~\cite{Watt:2018:MVW:3176245.3167082}.
These formal foundations are a crucial first step towards developing in-browser crypto
with guarantees of security.

In this paper, we go further, extending \wasm to become a verifiably secure
cryptographic language.
We augment \wasm's type system and semantics with cryptographically meaningful
types to produce Constant-Time WebAssembly (\ctwasm).
At the type level, \ctwasm allows developers to distinguish secret data (e.g.,
keys and messages) from public data.
This allows us to impose \textit{secure information flow}~\cite{Sabelfeld:2006:LIS} and \textit{constant-time} programming
disciplines~\cite{Barthe:2014:SNC:2660267.2660283,bearssl} on code that handles secret data and
ensure that well-typed \ctwasm code cannot leak such data, even via timing side channels.

\ctwasm brings together the convenience of in-browser \js crypto with the security
of a low-level, formally specified language.
\ctwasm allows application developers to incorporate third-party cryptographic
libraries of their choosing, much as they do today with \js.
But, unlike \js, \ctwasm ensures that these libraries
cannot leak secrets by construction---a property we guarantee via a fully
mechanized proof.

\ctwasm's type system draws from previous assembly language type systems
that enforce constant-time~\cite{Barthe:2014:SNC:2660267.2660283}.
Our system, however, is explicitly designed for the in-browser crypto use
case and is thus distinguished in two key ways.
First, like \wasm, we ensure that type checking is blisteringly fast, executing as a single linear pass.
Second, our type system makes trust relationships explicit: \ctwasm only allows
code explicitly marked as ``trusted'' to declassify data, bypassing the
security restrictions on secret data otherwise imposed by our type system.
%

\para{Contributions} In sum, this paper presents several contributions:

\begin{itemize}
\item \ctwasm: a new low-level bytecode language that extends \wasm with
cryptographically meaningful types to enable secure, in-browser crypto.
\item A fully mechanized formal model of the type system and operational
semantics of \ctwasm, together with a full proof of soundness, a verified type
checker, and proofs of several security properties, not least the \textit{constant-time} property (see Section \ref{sec:background:ct}).
\item Two implementations of \ctwasm: we extend the W3C specification reference
implementation and the real-world implementation of \wasm in V8.
\item Implementations, in \ctwasm, of several important cryptographic algorithms, including the TweetNaCl crypto library~\cite{tweetnacl}.
We experimentally evaluate our implementation work with respect to correctness, performance, and security.
\item Support tools that allow developers to \textbf{(1)} leverage the \ctwasm
  verifier to implement secure crypto code that will run on existing base \wasm
    implementations, in the style of the TypeScript compiler~\cite{typescript},
    and \textbf{(2)} semi-automatically infer \ctwasm annotations for base
    \wasm implementations.
\end{itemize}

\para{Open Source} All source and data are available under an open source
license at~\cite{supplement}.

\para{Paper Organization} We first review
WebAssembly and the constant-time programming paradigm
(Section~\ref{sec:background}) and give a brief overview of \ctwasm (Section~\ref{sec:overview}).
In Section~\ref{sec:semantics} we describe the \ctwasm language and its
semantics.
Our mechanized model and formal security guarantees are detailed in
Section~\ref{sec:model}.
We describe our implementations, supporting tools, and evaluation in
Sections~\ref{sec:implementation}.
We review related work in Section~\ref{sec:related}.
Finally we discuss future work in Section~\ref{sec:future} and conclude.


\section{Background}
\label{sec:background}

In this section we give a brief overview of the constant-time programming
paradigm and the WebAssembly bytecode language.
We then proceed to an overview of Constant-Time WebAssembly.

\subsection{Constant-time Programming Paradigm}
\label{sec:background:ct}
Naive implementations of cryptographic algorithms often leak
information---the very information they are designed to protect---via timing
side channels.
Kocher~\cite{kocher1996timing}, for example, shows how a textbook
implementation of RSA can be abused by an attacker to leak secret
key bits.
Similar key-recovery attacks were later demonstrated
on real implementations (e.g.,
RSA~\cite{brumley2005remote} and AES~\cite{bernstein2005cache, osvik2006cache}).
As a result, crypto-engineering best practices have shifted to mitigate such timing vulnerabilities.
%
Many modern cryptographic algorithms are even designed with such concerns from
the start~\cite{salsa20, poly1305, bernstein:curve25519}.

The prevailing approach for protecting crypto implementations against timing
attacks is to ensure that the code runs in ``constant time''.
An implementation is said to be \emph{constant-time} if its execution
time is not dependent on sensitive data, referred to as
\textit{secret} values (e.g., secret keys or messages).
Constant-time implementations ensure that an attacker observing their
execution behaviors cannot deduce any secret values.
Though the precise capabilities of attackers vary---e.g., an
attacker co-located with a victim has more capabilities than a remote attacker---most
secure crypto implementations follow a conservative
\emph{constant-time programming paradigm} that altogether avoids variable-time
operations, control flow, and memory access patterns that depend on
secrets~\cite{ctrules, bearssl}.

Verifying the constant-time property (or detecting lack thereof) for a given
implementation is considered one of the most important verification problems in
cryptography~\cite{ctverif,
Almeida:2016:VSS:3081631.3081644,
10.1007/978-3-319-66402-6_16,
jasmin, hacl, vale,
fact,
FiatCryptoSP19}.
To facilitate formal reasoning, these works typically represent this constant-time property using a
\textit{leakage model}~\cite{Boreale:2009:QIL:1539055.1539738} over a
small-step semantics for a given language.
A leakage model is a map from program state/action to an \textit{observation},
an abstract representation of an attacker's knowledge.
For each construct of the language, the leakage model encodes what information
is revealed (to an attacker) by its execution.
For example, the leakage model for branching operations such as \texttt{if} or
\texttt{while} leaks all values associated with the branch
condition, to represent that an attacker may use timing knowledge to reason about
which branch was taken~\cite{ctverif}.
Proving that a given program enjoys the constant-time property can then be abstracted
as a proof that the leakage accumulated over the course of the program's execution is
invariant with respect to the values of secret inputs.

In general, the leakage model of a system must
encompass the behavior of hardware and compiler optimizations across all
different platforms.
For example, for C, operators such as division and modulus, on some
architectures, are compiled to instruction sequences that have value-dependent
timing.
A conservative leakage model must accordingly encode
these operators as leaking the values of their operands~\cite{ctverif}.
While there is unavoidably a disconnect between the abstraction of a leakage
model and the actions of real-world compilers and architectures,
implementations that have such formal models have proven useful in practice.
For example, the HACL* library~\cite{hacl} has been adopted by
Firefox~\cite{hacl-ff}, while Fiat~\cite{FiatCryptoSP19} has been adopted by Chrome.

Unfortunately, much of this work does not translate well to the web platform.
Defining a leakage model for \js is extremely difficult.
\js has many complex language features that contribute to this
difficulty---from prototypes, proxies, to setters and
getters~\cite{ecmascript}.
Even if we restrict ourselves to well-behaving subsets of \js (e.g.,
asm.js~\cite{asmjs_spec} or defensive \js~\cite{djs}), the leakage model must
capture the behavior of \js runtimes---and their multiple just-in-time
compilers and garbage collectors.
 
Despite these theoretical shortcomings, \js crypto libraries remain overwhelmingly popular~\cite{signal-js,
kobeissi2017automated, npm-tweetnacl, npm-pbkdf2, npm-elliptic,
npm-crypto-browserify},
even in the presence of native libraries which were intended to curb their use in the
web ecosystem~\cite{halpin2014w3c}.
Unfortunately, these competing solutions proved inadequate. Native crypto libraries
differ wildly across platforms. For example the Web Crypto~\cite{halpin2014w3c} and the
Node.js crypto~\cite{node-crypto} APIs (available to browser and server-side \js respectively) barely overlap,
undercutting a major motivation for using \js in the first place---its cross-platform nature.
They are also unnecessarily complex (e.g., the Web Crypto API, like
OpenSSL, is ``\emph{the space shuttle of crypto
libraries}''~\cite{green:webcrypto}) when compared to state-of-the-art
libraries like NaCl~\cite{nacl}.
And, worst of all, none of these native libraries implement modern cryptographic
algorithms such as the Poly1305 Message Authentication Code~\cite{poly1305},
a default in modern crypto libraries~\cite{nacl}).
%
%
As we argue in this paper, WebAssembly can address these shortcomings,
in addition to those of \js. Because of its low-level nature, we can
sensibly relate existing work on assembly language leakage models to \wasm and provide
a formal, principled approach to reasoning about constant-time crypto code.
This gives us an excellent foundation on which to build \ctwasm.
Moreover, we will show that Poly1305, among many other cryptographic algorithms,
can be securely implemented in \ctwasm.
We next give an overview of \wasm and describe our extensions to the language.

\subsection{WebAssembly}
\
WebAssembly is a low-level bytecode language newly implemented by all major
browsers.
The stack-machine language is designed to allow developers to efficiently and
safely execute native code in the browser, without having to resort to
browser-specific solutions (e.g., Native Client~\cite{yee2009native}) or
subsets of \js (e.g., asm.js~\cite{asmjs_spec}).
Hence, while \wasm shares some similarities with low-level, assembly languages,
many \wasm design choices diverge from tradition.
We review three key design features relevant to writing secure crypto code:
\wasm's module system, type system, and structured programming paradigm.
We refer the reader to~\cite{Haas:2017:BWU:3062341.3062363} for an excellent,
in-depth overview of \wasm.

\para{Module system}
WebAssembly code is organized into \emph{modules}.
Each module contains a set of definitions: \emph{functions}, \emph{global
variables}, a \emph{linear memory}, and a \emph{table} of functions.
Modules are \emph{instantiated} by the embedding environment---namely
\js---which can invoke \wasm functions exported by the
module, manipulate the module's memory, etc.
%
At the same time, the embedding environment must also provide definitions
(e.g., from other \wasm modules) for functions the module declared as imports.
%

In a similar way to Safe Haskell~\cite{safehaskell}, we extend \wasm's module system
to further allow developers to specify if a particular import is trusted or
untrusted. In combination with our other type system extensions, this allows developers
to safely delineate the boundary between their own code and third-party,
untrusted code.

\para{Strong type system}
WebAssembly has a strong, static type system and an unambiguous formal
small-step semantics~\cite{Haas:2017:BWU:3062341.3062363}.
Together, these ensure that well-typed WebAssembly programs are ``safe'', i.e.,
they satisfy progress and preservation~\cite{Wright:1994:SAT:191905.191909, Watt:2018:MVW:3176245.3167082}.
This is especially important when executing \wasm code in the
browser---bytecode can be downloaded from arbitrary, potentially untrustworthy
parties.
Hence, before instantiating a module, \wasm engines validate (type check) the
module to ensure safety.
We extend the type system to enable developers to explicitly annotate secret
data and extend the type checker to ensure that secrets are not leaked
(directly or indirectly).

\para{Structured programming paradigm}
WebAssembly further differs from traditional assembly languages in
providing structured control flow constructs instead of simple
(direct/indirect) jump instructions.
Specifically, \wasm provides high-level control flow constructs for branching
(e.g., $\KK{if}$-$\KK{else}$ blocks) and looping (e.g., $\KK{loop}$ construct
with the $\KK{br\textunderscore{}if}$ conditional branch).
The structured control flow approach has many benefits.
For example, it ensures that \wasm code can be validated and compiled in a
single pass~\cite{Haas:2017:BWU:3062341.3062363}.
This provides a strong foundation for our extension: we can enforce a constant-time
leakage model via type checking, in a single pass, instead of a more complex static
analysis~\cite{Barthe:2014:SNC:2660267.2660283}.

These design features position WebAssembly as an especially good language to
extend with a light-weight information flow type system that can automatically
impose the constant-time discipline on crypto code~\cite{Volpano:1996:STS:353629.353648,
Sabelfeld:2006:LIS,Barthe:2014:SNC:2660267.2660283}.
In the next section, we give an overview of our extension: \ctwasm.


\section{Constant-Time WebAssembly, an overview}
\label{sec:overview}

We extend \wasm to enable developers to implement cryptographic
algorithms that are verifiably constant-time.
Our extension, Constant-Time WebAssembly, is rooted in three main design principles.
First, \ctwasm should allow developers to explicitly specify the sensitivity of
data and automatically ensure that code handling secret data cannot leak the
data.
To this end, we extend the \wasm language with new secret values (e.g., secret
32-bit integers typed $\K{s32}$) and secret memories.
We also extend the type system of \wasm to ensure that such
secret data cannot be leaked either directly (e.g., by writing secret values to public
memory) or indirectly (e.g., via control flow and memory access patterns), by
imposing secure information flow and constant-time disciplines on code that handles secrets
(see Section~\ref{sec:semantics}).
We are careful to design \ctwasm as a strict syntactic and semantic superset of
\wasm---all existing \wasm code is valid \ctwasm---although no security benefits
are guaranteed without additional secrecy annotations.

Second, since \wasm and most crypto algorithms are designed with performance in
mind, \ctwasm must not incur significant overhead, either from validation or execution.
This is especially true for the web use case.  Overall page load time is considered
one of---if not \textit{the}---key website performance
metric~\cite{mozillaloadtime,maxcdnloadtime}, so requiring the web client to
conduct expensive analyses of loaded code before execution would be infeasible.
Our type-driven approach addresses this design goal---imposing almost no
runtime overhead.
%
%
\ctwasm only inserts dynamic checks for indirect function calls via
$\KK{call\textunderscore{}indirect}$; much like \wasm itself, this ensures that
types are preserved even for code that relies on dynamic dispatch.
In contrast to previous type checking algorithms for constant-time low-level
code~\cite{Barthe:2014:SNC:2660267.2660283, ctverif}, \ctwasm leverages
\wasm's structured control flow and strongly-typed design to implement an
efficient type checking algorithm---in a single pass, we can verify if a piece
of crypto code is information flow secure and constant-time
(see Section~\ref{sec:security}).
We implement this type checker in the V8 engine and as a standalone verified
tool.
Our standalone type checker can be used by crypto engineers during development
to ensure their code is constant-time, even when ``compiled'' to legacy \wasm
engines without our security annotations (see Section~\ref{sec:implementation:tools:strip}).

Third, \ctwasm should be flexible enough to implement real-world crypto
algorithms.
To enforce constant-time programming, our type system is more restrictive than
more traditional information flow control type systems (e.g.,
JIF's~\cite{myers2001jif, myers1999jflow} or FlowCaml's~\cite{flowcaml}).
For example, we do not allow allow branching (e.g., via
$\KK{br\textunderscore{}if}$ or $\KK{loop}$) on secret data or secret-depended
memory instructions ($\KK{load}$s and $\KK{store}$s).
These restrictions, however, are no more onerous than what developers already
impose upon themselves~\cite{ctrules, bearssl}: our type checker effectively
ensures that $\K{untrusted}$ code respects these (previously) self-imposed limitations.

\begin{tikzpicture}[remember picture,overlay]
\node[above left,inner sep=0pt] at (current page text area.south east) {%
\begin{minipage}{0.5\textwidth}
 \raggedright
\begin{adjustwidth}{2em}{0pt}
(\KK{func} \$nacl\_secretbox\_open\\
\begin{adjustwidth}{2.5em}{0pt}
...\\
(\KK{call} \$crypto\_onetimeauth\_verify)\\
(\KK{i32.declassify})\\
(\KK{i32.const} \K{0})\\
(\KK{if} (\KK{i32.eq}) (\KK{then}\\
\begin{adjustwidth}{2.5em}{0pt}
...\\
(\KK{call} \$crypto\_stream\_xor)\\
(\KK{return} (\KK{i32.const} \K{0}))))\\
\end{adjustwidth}
(\KK{return} (\KK{i32.const} \K{-1})))
\end{adjustwidth}
\end{adjustwidth}
  \captionof{figure}{Verified decryption in TweetNaCl relies on a declassification to
  terminate early (if verification fails).}
\label{fig:declass}
\end{minipage}};
\end{tikzpicture}
\parshape=7 
0pt \linewidth 
0pt \linewidth 
0pt \linewidth 
0pt \linewidth 
0pt \linewidth 
0pt \linewidth 
0pt \dimexpr0.45\linewidth\relax 
\ctwasm does, however, provide an escape hatch that
allows developers to bypass our strict requirements: explicit declassification
with a new $\KK{declassify}$ instruction, which can be used to
downgrade the sensitivity of data from \K{secret} to \K{public}.
This is especially useful when developers encrypt data and thus no longer need the
type system to protect it or, as Fig.~\hyperlink{fig-declass-para}{1} shows, to reveal
(and, by choice, explicitly leak) the success or failure of a verification
algorithm.
\ctwasm allows these use cases, but makes them explicit in the code.
\hypertarget{fig-declass-para}
To ensure declassification is not abused, our type system restricts the
use of $\KK{declassify}$ to functions marked as $\K{trusted}$.
This notion of trust is transitively enforced across function and module
boundaries: functions that call $\K{trusted}$ functions must themselves be
marked $\K{trusted}$.
This ensures that developers cannot accidentally leak secret data without
explicitly opting to trust such functions (e.g., when declaring module
imports).
Equally important, by marking a function $\K{untrusted}$, developers give a swiftly
verifiable contract that its execution cannot leak secret
data directly, indirectly via timing channels, or via declassification.

\para{Trust model}
Our attacker model is largely standard.
We assume an attacker that can
\textbf{(1)} supply and execute arbitrary $\K{untrusted}$ \ctwasm functions on
secret data and
\textbf{(2)} observe the runtime behavior of this code, according to the
leakage model we define in Section~\ref{sec:model}.
Since \ctwasm does not run in isolation, we assume that the \js embedding
environment and all $\K{trusted}$ \ctwasm functions are correct and cannot be
abused by the attacker to leak sensitive data.
Under this model, \ctwasm guarantees that the attacker will not learn any
secrets.
In practice, these guarantees allow application developers to execute
untrusted, third-party crypto libraries (e.g., from content distribution
networks or package managers such as NPM) without fear that leakage will occur.

\section{\ctwasm Semantics}
\label{sec:semantics}

\begin{figure*}
\fontsize{8pt}{8pt}
$$
\begin{array}{@{}l@{}l@{}}
\begin{array}{@{}l@{}}
\begin{array}{@{}l@{~}r@{~}c@{~}l@{}}
\text{(immediates)} & \x{imm} &::=& \x{nat} \\
\hilite{\text{(secrecy types)}} & \hilite{\x{sec}} & \hilite{::=}&
  \hilite{\K{secret} ~|~
  \K{public}}  \\
\hilite{\text{(trust types)}} & \hilite{tr} & \hilite{::=} &
  \hilite{\K{trusted} ~|~
  \K{untrusted}}  \\
\text{(packed types)} & \x{pt} &::=&
  \K{i8} ~|~
  \K{i16} ~|~
  \K{i32} \\
\text{(value types)} & t &::=&
  \K{i32}\hilite{'~\x{sec}} ~|~
  \K{i64}\hilite{'~\x{sec}} ~|~
  \K{f32} ~|~
  \K{f64} \\
\text{(function types)} & \x{ft} &::=&
  t^\ast \Kto t^\ast \\
\text{(global types)} & \x{gt} &::=&
  \K{mut}^?~t \\
\end{array}
\\~\vspace{-2ex}\\
\begin{array}{@{}l@{~}r@{~}c@{~}l@{}}
\multicolumn{2}{@{}r}{\x{unop}_{\K{i}N}} & ::= &
  \KK{clz} ~|~
  \KK{ctz} ~|~
  \KK{popcnt} \\
\multicolumn{2}{@{}r}{\x{unop}_{\K{f}N}} & ::= &
  \KK{neg} ~|~
  \KK{abs} ~|~
  \KK{ceil} ~|~
  \KK{floor} ~|~ \\&&&
  \KK{trunc} ~|~
  \KK{nearest} ~|~
  \KK{sqrt} \\
\multicolumn{2}{@{}r}{\x{binop}_{\K{i}N}} & ::= &
  \KK{add} ~|~
  \KK{sub} ~|~
  \KK{mul} ~|~
  \KK{div\textunderscore}\x{sx} ~|~ \\&&&
  \KK{rem\textunderscore}\x{sx} ~|~
  \KK{and} ~|~
  \KK{or} ~|~
  \KK{xor} ~|~ \\&&&
  \KK{shl} ~|~
  \KK{shr\textunderscore}\x{sx} ~|~
  \KK{rotl} ~|~
  \KK{rotr} \\
\multicolumn{2}{@{}r}{\x{binop}_{\K{f}N}} & ::= &
  \KK{add} ~|~
  \KK{sub} ~|~
  \KK{mul} ~|~
  \KK{div} ~|~ \\&&&
  \KK{min} ~|~
  \KK{max} ~|~
  \KK{copysign} \\
\multicolumn{2}{@{}r}{\x{testop}_{\K{i}N}} & ::= &
  \KK{eqz} \\
\multicolumn{2}{@{}r}{\x{relop}_{\K{i}N}} & ::= &
  \KK{eq} ~|~
  \KK{ne} ~|~
  \KK{lt\textunderscore}\x{sx} ~|~
  \KK{gt\textunderscore}\x{sx} ~|~ \\&&&
  \KK{le\textunderscore}\x{sx} ~|~
  \KK{ge\textunderscore}\x{sx} \\
\multicolumn{2}{@{}r}{\x{relop}_{\K{f}N}} & ::= &
  \KK{eq} ~|~
  \KK{ne} ~|~
  \KK{lt} ~|~
  \KK{gt} ~|~
  \KK{le} ~|~
  \KK{ge} \\
\multicolumn{2}{@{}r}{\x{cvtop}} & ::= &
  \KK{convert} ~|~
  \KK{reinterpret} ~|~ \\&&&
  \hilite{\KK{classify} ~|~
  \KK{declassify}} \\
\multicolumn{2}{@{}r}{\x{sx}} & ::= &
  \KK{s} ~|~
  \KK{u} \\
\end{array}
\end{array}
\hspace{-3ex}
\begin{array}{@{}l@{}}
\begin{array}{@{}l@{~~}r@{~}c@{~}l@{}}
\text{(constants)} & k &::=&
  \dots \\
\text{(instructions)} & e &::=&
  \KK{unreachable} ~|~
  \KK{nop} ~|~
  \KK{drop} ~|~
  \KK{select}~\hilite{\x{sec}}~|~ \\&&&
  \KK{block}~\x{ft}~e^\ast~\KK{end} ~|~
  \KK{loop}~\x{ft}~e^\ast~\KK{end} ~|~ \\&&&
  \KK{if}~\x{ft}~e^\ast~\KK{else}~e^\ast~\KK{end} ~|~
  t\KK{.const}~k ~|~ \\&&&
  \KK{br}~\x{imm} ~|~
  \KK{br\textunderscore{}if}~\x{imm} ~|~
  \KK{br\textunderscore{}table}~\x{imm}^+ ~|~ \\&&&
  \KK{return} ~|~
  \KK{call}~\x{imm} ~|~
  \KK{call\textunderscore{}indirect}~(\hilite{\x{tr}},\x{ft}) ~|~ \\&&&
  \KK{get\textunderscore{}local}~\x{imm} ~|~
  \KK{set\textunderscore{}local}~\x{imm} ~|~ \\&&&
  \KK{tee\textunderscore{}local}~\x{imm} ~|~
  \KK{get\textunderscore{}global}~\x{imm} ~|~ \\&&&
  \KK{set\textunderscore{}global}~\x{imm} ~|~ \\&&&
  t\KK{.load}~(\x{pt}\KK{\textunderscore}\x{sx})^?~a~o ~|~
  t\KK{.store}~\x{pt}^?~a~o ~|~ \\&&&
  \KK{memory.size} ~|~
  \KK{memory.grow} ~|~ \\&&&
  t\KK{.}\x{unop}_t ~|~
  t\KK{.}\x{binop}_t ~|~
  t\KK{.}\x{testop}_t ~|~ \\&&&
  t\KK{.}\x{relop}_t ~|~
  t\KK{.}\x{cvtop}~t\KK{\textunderscore}\x{sx}^? \\
\end{array}
\\~\\[-1ex]\hspace{0ex}
\begin{array}{@{}l@{~}r@{~}c@{~}l@{}}
\text{(functions)} & \x{func} &::=&
  \x{ex}^\ast~\KK{func}~(\hilite{\x{tr}},\x{ft})~\KK{local}~\x{t}^\ast~e^\ast ~|~ \\&&&
  \x{ex}^\ast~\KK{func}~(\hilite{\x{tr}},\x{ft})~\x{imp} \\
\text{(globals)} & \x{glob} &::=&
  \x{ex}^\ast~\KK{global}~\x{gt}~e^\ast ~|~
  \x{ex}^\ast~\KK{global}~\x{gt}~\x{imp} \\
\text{(tables)} & \x{tab} &::=&
  \x{ex}^\ast~\KK{table}~\x{n}~\x{imm}^\ast ~|~
  \x{ex}^\ast~\KK{table}~\x{n}~\x{imp} \\
\text{(memories)} & \x{mem} &::=&
  \x{ex}^\ast~\KK{memory}~\x{n}~\hilite{\x{sec}}~|~
  \x{ex}^\ast~\KK{memory}~\x{n}~\hilite{\x{sec}}~\x{imp} \\
\text{(imports)} & \x{imp} &::=&
  \KK{import}~\text{``\x{name}''}~\text{``\x{name}''} \\
\text{(exports)} & \x{ex} &::=&
  \KK{export}~\text{``\x{name}''} \\
\text{(modules)} & \x{mod} &::=&
  \KK{module}~\x{func}^\ast~\x{glob}^\ast~\x{tab}^?~\x{mem}^? \\ 
\end{array}
\\~\\[-1ex]\hspace{0ex}
\hilite{
\begin{array}{@{}l@{~~}c@{~~}l @{\hspace{1.2cm}}l@{~~}c@{~~}l}
\K{sec}~(\K{i}N'~\x{sec}) & \triangleq & \x{sec} & \K{i}N & ::= & \K{i}N'~\K{public}\\
\K{sec}~\K{f}N & \triangleq & \K{public}  & \K{s}N & ::= & \K{i}N'~\K{secret}\\
\end{array}
}
\end{array}
\end{array}
$$


\caption{\ctwasm abstract syntax as an extension \hilite{(highlighted) } of the grammar given by~\cite{Haas:2017:BWU:3062341.3062363}.}
\label{fig:syntax}
\vspace{-1\baselineskip}
\end{figure*}


\begin{figure*}
\footnotesize\fontsize{9pt}{9pt}

$$
\text{(contexts) }~C ::= \Big\{
\begin{array}{l}
  \hilite{{\K{trust}} \; \x{tr}},\; {\K{func}} \; (\hilite{\x{tr}},\x{ft})^\ast,\; {\K{global}} \; \x{gt}^\ast,\; {\K{table}} \; \x{n}^?,\;\\
{\K{memory}} \; (\x{n},\hilite{\x{sec}})^?,\; {\K{local}} \; t^\ast,\; \void{{\K{return}} \; t^\ast,\;} {\K{label}} \; (t^\ast)^\ast,\; {\K{return}} \; (t^\ast)^?
\end{array}
\Big\}
\vspace{2ex}
$$

\hilite{$tr \succtr tr' \triangleq (tr = tr') \lor (tr = \texttt{trusted} \land tr' = \texttt{untrusted})$}

\vspace{-0ex}


$$
\frac{
}{
  C \vdash t\KK{.const}~c : \epsilon \to t
}
\qquad
\frac{
}{
  C \vdash t\KK{.}\x{unop} : t \to t
}
\qquad
\frac{
}{
  C \vdash t\KK{.}\x{binop} : t~t \to t
}
$$

$$
\frac{
  \hilite{\K{sec}~\x{t} = \x{sec}}
}{
  C \vdash t\KK{.}\x{testop} : t \to (\K{i32}\hilite{'~\x{sec}})
}
\qquad
\frac{
  \hilite{\K{sec}~\x{t} = \x{sec}}
}{
  C \vdash t\KK{.}\x{relop} : t~t \to (\K{i32}\hilite{'~\x{sec}})
}
$$

$$
\frac{
  t_1 \neq t_2
  \qquad
  \x{sx}^? = \epsilon \Leftrightarrow (t_1 = \KK{i}n_1'~\hilite{\x{sec}} \wedge t_2 = \KK{i}n_2'~\hilite{\x{sec}} \wedge|t_1| < |t_2|) \vee (t_1 = \KK{f}n \wedge t_2 = \KK{f}n')
}{
  C \vdash t_1\KK{.convert}~t_2\K{\textunderscore}\x{sx}^? : t_2 \to t_1
}
$$

$$
\frac{
  t_1 \neq t_2
  \qquad
  |t_1| = |t_2|
  \qquad
  \hilite{\K{sec}~t_1 = \K{sec}~t_2}
}{
  C \vdash t_1\KK{.reinterpret}~t_2 : t_2 \to t_1
}
\qquad
\hilite{
\frac{
(t_1 = \KK{i}n'~\K{secret} \wedge t_2 = \KK{i}n'~\K{public})
}{
  C \vdash t_1\KK{.classify}~t_2 : t_2 \to t_1
}%
}%
\vspace{-1em}
$$

$$
\hilite{
\frac{
  C_{\K{trust}} = \K{trusted}
  \qquad
(t_1 = \KK{i}n'~\K{public} \wedge t_2 = \KK{i}n'~\K{secret})
}{
  C \vdash t_1\KK{.declassify}~t_2 : t_2 \to t_1
}%
}%
$$
$$
\frac{
}{
  C \vdash \KK{unreachable} : t_1^\ast \to t_2^\ast
}
\quad
\frac{
}{
  C \vdash \KK{nop} : \epsilon \to \epsilon
}
\quad
\frac{
}{
  C \vdash \KK{drop} : t \to \epsilon
}
\quad
\frac{
\hilite{~\x{sec} = \K{secret} \longrightarrow ~\K{sec}~\x{t} = \K{secret}}
}{
  C \vdash \KK{select}~\hilite{\x{sec}} : t~t~(\K{i32}\hilite{'~\x{sec}}) \to t
}
$$

$$
\frac{
  \x{ft} = t_1^n \to t_2^m
  \qquad
  C,{{\K{label}}}\,(t_2^m) \vdash e^\ast : \x{ft}
}{
  C \vdash \KK{block}~\x{ft}~e^\ast~\KK{end} : \x{ft}
}
\qquad
\frac{
  \x{ft} = t_1^n \to t_2^m
  \qquad
  C,{{\K{label}}}\,(t_1^n) \vdash e^\ast : \x{ft}
}{
  C \vdash \KK{loop}~\x{ft}~e^\ast~\KK{end} : \x{ft}
}
$$

$$
\frac{
  \x{ft} = t_1^n \to t_2^m
  \qquad
  C,{{\K{label}}}\,(t_2^m) \vdash e_1^\ast : \x{ft}
  \qquad
  C,{{\K{label}}}\,(t_2^m) \vdash e_2^\ast : \x{ft}
}{
  C \vdash \KK{if}~\x{ft}~e_1^\ast~\KK{else}~e_2^\ast~\KK{end} : t_1^n~\K{i32} \to t_2^m
}
\qquad
\frac{
  C_{\K{return}} = t^\ast
}{
  C \vdash \KK{return} : t_1^\ast~t^\ast \to t_2^\ast
}
$$

$$
\frac{
  C_{\K{label}}(i) = t^\ast
}{
  C \vdash \KK{br}~i : t_1^\ast~t^\ast \to t_2^\ast
}
\qquad
\frac{
  C_{\K{label}}(i) = t^\ast
}{
  C \vdash \KK{br\textunderscore{}if}~i : t^\ast~\K{i32} \to t^\ast
}
\qquad
\frac{
  (C_{\K{label}}(i) = t^\ast)^+
}{
  C \vdash \KK{br\textunderscore{}table}~i^+ : t_1^\ast~t^\ast~\K{i32} \to t_2^\ast
}
$$

$$
\frac{
  \hilite{C_{\K{trust}} = \x{tr}}
  \quad
  C_{\K{func}}(i) = (\hilite{\x{tr'}},\x{ft})
  \quad
  \hilite{tr \succtr tr'}
}{
  C \vdash \KK{call}~i : \x{ft}
}
\qquad
\frac{
  \x{ft} = t_1^\ast \to t_2^\ast
  \qquad
  \hilite{C_{\K{trust}} = \x{tr}}
  \qquad
  \hilite{tr \succtr tr'}
  \qquad
  C_{\K{table}} = \x{n}
}{
  C \vdash \KK{call\textunderscore{}indirect}~(\hilite{\x{tr'}},\x{ft}) : t_1^\ast~\K{i32} \to t_2^\ast
}
$$

$$
\frac{
  C_{\K{local}}(i) = t
}{
  C \vdash \KK{get\textunderscore{}local}~i : \epsilon \to t
}
\qquad
\frac{
  C_{\K{local}}(i) = t
}{
  C \vdash \KK{set\textunderscore{}local}~i : t \to \epsilon
}
\qquad
\frac{
  C_{\K{local}}(i) = t
}{
  C \vdash \KK{tee\textunderscore{}local}~i : t \to t
}
$$

$$
\frac{
  C_{\K{global}}(i) = \K{mut}^?~t
}{
  C \vdash \KK{get\textunderscore{}global}~i : \epsilon \to t
}
\qquad
\frac{
  C_{\K{global}}(i) = \K{mut}~t
}{
  C \vdash \KK{set\textunderscore{}global}~i : t \to \epsilon
}
$$

$$
\frac{
  C_{\K{memory}} = (\x{n},\hilite{\x{sec}})
  \qquad
  \hilite{\K{sec}~\x{t} = \x{sec}}
  \qquad
  2^a \leq (|\x{tp}| <)^? |t|
  \qquad
  (\x{tp}\K\textunderscore\x{sz})^? = \epsilon \vee t = \KK{i}m\hilite{'~\x{sec}}
}{
  C \vdash t\KK{.load}~(\x{tp}\K\textunderscore\x{sz})^?~a~o : \K{i32} \to t
}
$$

$$
\frac{
  C_{\K{memory}} = (\x{n},\hilite{\x{sec}})
  \qquad
  \hilite{\K{sec}~\x{t} = \x{sec}}
  \qquad
  2^a \leq (|\x{tp}| <)^? |t|
  \qquad
  \x{tp}^? = \epsilon \vee t = \KK{i}m\hilite{'~\x{sec}}
}{
  C \vdash t\KK{.store}~\x{tp}^?~a~o : \K{i32}~t \to \epsilon
}
$$

$$
\frac{
  C_{\K{memory}} = (\x{n},\hilite{\x{sec}})
}{
  C \vdash \KK{memory.size} : \epsilon \to \K{i32}
}
\qquad
\frac{
  C_{\K{memory}} = (\x{n},\hilite{\x{sec}})
}{
  C \vdash \KK{memory.grow} : \K{i32} \to \K{i32}
}
$$

$$
\frac{
}{
  C \vdash \epsilon : \epsilon \to \epsilon
}
\qquad
\frac{
  C \vdash e_1^\ast : t_1^\ast \to t_2^\ast
  \qquad
  C \vdash e_2 : t_2^\ast \to t_3^\ast
}{
  C \vdash e_1^\ast~e_2 : t_1^\ast \to t_3^\ast
}
\qquad
\frac{
  C \vdash e^\ast : t_1^\ast \to t_2^\ast
}{
  C \vdash e^\ast : t^\ast~t_1^\ast \to t^\ast~t_2^\ast
}
$$

\caption{\ctwasm typing rules as an extension \hilite{(highlighted) } of the typing rules given by~\cite{Haas:2017:BWU:3062341.3062363}.}
\label{fig:typing}

\end{figure*}


\begin{figure*}
\fontsize{8pt}{8pt}
$$
\begin{array}{@{}ll@{}}
\begin{array}[b]{@{}llcl@{}}
\text{(values)} & \x{v} &::=&
  t\KK{.const}~k \\
\text{(store index)} & \x{a} &::=&
  \x{imm} \\
\text{(module instances)} & \x{inst} &::=&
  \{{\K{func\_i}} \; \x{a}^\ast,\; {\K{global\_i}} \; \x{a}^\ast,\; {\K{table\_i}} \; \x{a}^?,\; {\K{mem\_i}} \; \x{a}^?\} \\
\text{(function closures)} & \x{cl} &::=&
  \{{\K{instance\_ind}} \; \x{a},\; \K{type}~(\hilite{\x{tr}},\x{ft}), \; {\K{code}} \; \x{func}\} ~|~
  \{{\K{type}~(\hilite{\x{tr}},\x{ft}), \K{host}} \; ...\} \\
\text{(memory instances)} & \x{mi} &::=&
  \x{byte}^\ast \\%
\text{(store)} & s &::=&
  \{{\K{inst}} \; \x{inst}^\ast,\; {\K{func}} \; \x{cl}^\ast,\; {\K{global}} \; (\K{mut}^?~\x{v})^\ast,\; {\K{table}} \; \x{cl}^\ast,\; {\K{mem} \; (\hilite{\x{sec}}, \x{mi})^\ast}\} \\[1ex]

\text{(administrative instructions)} & e &::=&
  \dots ~|~
  \KK{trap} ~|~
  \KK{callcl}~\x{cl} ~|~
  \KK{label}_n\!\{e^\ast\}~e^\ast~\KK{end} ~|~
  \KK{local}_n\!\{i; \x{v}^\ast\}~e^\ast~\KK{end}\\


\text{(configurations)} & \x{c} &::=&
  \x{s}; \x{v}^\ast; \x{e}^\ast \\[1ex]
\end{array}
\end{array}
$$


$$
\fontsize{9pt}{9pt}
\begin{array}{@{}rcl@{}}
\hilite{s; \x{vs};~(\K{s}N.\KK{const}~k)~t_2\KK{.declassify}~t_1}&\hilite{\leadsto_i}&
  \hilite{s; \x{vs};~(\K{i}N.\KK{const}~k)} \\
\hilite{s; \x{vs};~(\K{i}N.\KK{const}~k)~t_2\KK{.classify}~t_1}&\hilite{\leadsto_i}&
  \hilite{s; \x{vs};~(\K{s}N.\KK{const}~k)} \\
s; \x{vs};~\x{v}_1~\x{v}_2~((\KK{i32}\hilite{'~\x{sec}})\KK{.const}~0)~\KK{select}~\hilite{\x{sec}'} &\leadsto_i&
 s; \x{vs}; \x{v}_2 \\
s; \x{vs};~\x{v}_1~\x{v}_2~((\KK{i32}\hilite{'~\x{sec}})\KK{.const}~k+1)~\KK{select}~\hilite{\x{sec}'} &\leadsto_i&
  s; \x{vs};\x{v}_1 \\
\vspace{0.1cm}
s; \x{vs};~(\KK{i32}\KK{.const}~k)~\KK{call\_indirect}~(\hilite{\x{tr}},\x{ft}) &\leadsto_i&
  s; \x{vs};\KK{callcl}~\x{cl} \qquad \textnormal{\pbox{0.3\textwidth}{if \K{table\_i} ((\K{inst} $s$)!$i$) = $a$ \\ and ((\K{table}~i)!$a$)!$k$ = $cl$ \\ and \K{type} ~\x{cl} = $(\hilite{\x{tr}},\x{ft})$ }} \quad (*)\\
s; \x{vs};~(\KK{i32}\KK{.const}~k)~\KK{call\_indirect}~(\hilite{\x{tr}},\x{ft}) &\leadsto_i&
  s; \x{vs};\KK{trap} \qquad \quad \hspace{0.1cm} \text{otherwise} \\
[1ex]
\end{array}
$$

$$
\fontsize{9pt}{9pt}
\begin{array}{@{}ll@{}}
(*) & \pbox{0.75\textwidth}{\textnormal{\KK{callcl}~\x{cl} represents a function closure about to be entered as a local context. It is used to define a unifying dynamic semantics for the various forms of function call in \wasm, and its semantics is unchanged from~\cite{Haas:2017:BWU:3062341.3062363}.}}
\end{array}
$$


\caption{Selected \ctwasm semantic definitions, extended \hilite{(highlighted) } from~\cite{Haas:2017:BWU:3062341.3062363}.
  Since the vast majority of the reduction rules are unchanged, we give only a few examples here.}
\label{fig:reduction}

\end{figure*}

We specify \ctwasm primarily as an extension of WebAssembly's syntax and type
system, with only minor extensions to its dynamic semantics.
Our new secrecy and trust annotations are designed to track the flow of secret
values through the program and restrict their usage to ensure both secure
information flow and constant-time security.
We give the \ctwasm extended syntax in Fig.~\ref{fig:syntax}, the
core type system in Fig.~\ref{fig:typing}, and
an illustrative selection of the runtime reduction rules in Fig.~\ref{fig:reduction}.

We now consider aspects of the base WebAssembly
specification, and describe how they are extended to form Constant-Time WebAssembly.


\subsection{Instances}

WebAssembly's typing and runtime execution are defined with respect to
a module \emph{instance}. An instance is a representation of the global state
accessible to a WebAssembly configuration (program) from link-time onwards.
In Fig.~\ref{fig:typing}, the typing context $C$ abstracts the
\emph{current instance}.
In Fig.~\ref{fig:reduction}, the small-step runtime reduction relation is indexed by
the current instance $i$.

Instances are effectively a collection of indexes into the \emph{store}, which
keeps track of all global state potentially shared between different configurations.\footnote{
  In ~\cite{Haas:2017:BWU:3062341.3062363},
  all instances are held as a list in the store, with evaluation rules
  parameterized by an index into this list.
  The ``live'' specification recently changed this so that evaluation rules are directly parameterized
  by an instance~\cite{instancespec}.
  We give our semantics as an extension of the original paper definition,
  although the transformation is ultimately trivial, so we will often refer to
  the current instance index as the ``current instance''.
}
If an element of the WebAssembly store (e.g., another module's memory or
function) is not indexed by the current instance, the executing WebAssembly
code is, by construction, prevented from accessing it.
%
%
%

\subsection{Typing and Value Types}
\wasm is a stack-based language.
Its primitive operations produce and consume a fixed number of \textit{value types}.
\wasm's type system assigns each operation a type of the form \mbox{$t^\ast \to {t'}^\ast$},
describing (intuitively) that the operation requires a stack of values of type $t^\ast$ to execute, and will produce a stack of values of type ${t'}^\ast$ upon completion.
The type of a \wasm code section (a list of operations) is the composition of these types, with a given operation potentially consuming the results of previous operations.

Base \wasm has four value types: $\K{i32}$, $\K{i64}$, $\K{f32}$, and
$\K{f64}$, representing 32 and 64 bit integer and floating point values,
respectively.
To allow developers to distinguish between public and secret data,
we introduce new value types which denote \textit{secret values}.
Formally, we first define secrecy annotations, \textit{sec}, which can take two
possible values: $\K{secret}$ or $\K{public}$.
We then extend the integer value types so that they are parameterized by this
annotation.
For syntactic convenience, we define the existing $\K{i32}$ and $\K{i64}$
WebAssembly type annotations as denoting \textit{public (integer) values}, with
new annotations $\K{s32}$ and $\K{s64}$ representing \textit{secret (integer)
values}.
Floating point types are always considered $\K{public}$, since most floating
point operations are variable-time and vulnerable to timing
attacks~\cite{Andrysco:2015:SFP:2867539.2867676, kohlbrenner2017, ctfp}.
 
As shown in Fig.~\ref{fig:typing}, all \ctwasm instructions (except
$\KK{declassify}$) preserve the secrecy of data.
We do not introduce any subtyping or polymorphism of secrecy for
existing \wasm operations over integer values;
pure WebAssembly seeks to avoid polymorphism in its type system wherever
practical, a paradigm we continue to emulate.
Instead, we make any necessary conversions explicit in the syntax of \ctwasm.
For example, the existing $\K{i32.}\KK{add}$ instruction of \wasm is
interpreted as operating over purely $\K{public}$ integers, while a new
$\K{s32.}\KK{add}$ instruction is added for $\K{secret}$ integers.
We introduce an explicit $\KK{classify}$ operation which relabels a public
integer value as a secret.
This allows us to use $\K{public}$ values wherever $\K{secret}$ values are
required; this is safe, and makes such a use explicit in the representation of
the program.

Together with the control flow and memory access restrictions described below,
our type system guarantees an information flow property: ensuring that, except
through $\KK{declassify}$, $\K{public}$ computations can never depend on
$\K{secret}$ values.
We give a mechanized proof of this in Section~\ref{sec:security}.

\subsection{Structured Control Flow}

Our type system enforces a constant-time discipline on $\K{secret}$ values.
This means that we do not allow secret values to be used as conditionals in
control flow instructions, such as \KK{call\_indirect}, \KK{br\_if},
\KK{br\_table}, or \KK{if}; only public values can be used as conditionals.
This is an onerous restriction, but it is one that cryptography implementers
habitually inflict on themselves in pursuit of security.
Indeed, it is described as best-practice in cryptography implementation style
guides~\cite{ctrules}, and as discussed throughout this paper, many theoretical
works on constant-time model such operations as unavoidably leaking the value
of the conditional to the attacker.

Our type system does, however, allow for a limited form of secret conditionals
with the \KK{select} instruction.
This instruction takes three operands and returns the first or second depending on the
third, condition operand.
Since secrecy of the conditional can be checked statically by the type system, secrecy
annotations have no effect on the dynamic semantics of Fig.~\ref{fig:reduction}.
Importantly, \KK{select} can do this without branching: conditional move
instructions allow \KK{select} to be implemented using a single, constant-time
hardware instruction~\cite{guide2016intel} and, for processors without such instructions
a multi-instruction arithmetic solution exists~\cite{ctrules}.
In either case, to preserve the constant-time property if the conditional is \K{secret}, both arguments to \KK{select} must
be fully evaluated. This is the case in the \wasm abstract machine, but real engines must
ensure that they respect this when implementing optimizations.
We extend the \KK{select} instruction with a secrecy annotation; a \KK{select} \K{secret} instruction
preserves constant-time (permitting \K{secret} conditionals), but may permit fewer optimizations.

\subsection{Memory}

Though $\K{secret}$ value types allow us to track the secrecy of stack values,
this is not enough.
\wasm also features linear memories, which can also be used to store and load
values.
We thus annotate each linear memory with \textit{sec}.
Our type system ensures that \K{public} (resp. \K{secret}) values can only be
stored in memories annotated \K{public} (resp. \K{secret}).
Dually, it ensures that loads from memory annotated \textit{sec} can only
produce \textit{sec} values.
To ensure that accessing memory does not leak any
information~\cite{osvik2006cache, brumley2005remote}, our type system also
require that all memory indices to $\KK{load}$ and $\KK{store}$ be \K{public}.
These restrictions preserve our information flow requirements in the presence
of arbitrary memory operations.

Our coarse-grained approach to annotating memory is not without trade offs.
Since \wasm only allows one memory per module, to store
both \K{public} and \K{secret} data in memory, a developer must create a second module
and import accessor functions for that module's memory.
%
A simple micro benchmark implementing this pattern reveals a 30\%
slowdown for the memory operations.
In practice, this is not a huge concern.
Once base \wasm gains support for multiple memories~\cite{wasm_ref_types}, a
module in \ctwasm could have both \K{public} and \K{secret} memories;
we choose not to implement our own solution now so as to maintain forwards
compatibility with the proposed \wasm extension.
Moreover, as we find in our evaluation (Section~\ref{sec:implementation:eval}),
many crypto algorithms don't require both \K{secret} and \K{public} memory in
practice.

A yet more sophisticated and fine-grained design would annotate individual
memory cells.
We eschew this design largely because it would demand a more complex (and thus
slower) type-checking algorithm (e.g., to ensure that a memory access at a
dynamic offset is indeed of the correct sensitivity).

\subsection{Trust and Declassification}

As previously mentioned, it is sometimes necessary for \ctwasm to allow
developers to bypass the above restrictions and cast $\K{secret}$ values to
$\K{public}$.
For example, when implementing an encryption algorithm, there is a point where
we transfer trust away from information flow security to the computational
hardness of cryptography.
At this point, the $\K{secret}$ data can be \emph{declassified} to $\K{public}$
(e.g., to be ``leaked'' to the outside world).

As a dual to $\KK{classify}$ we provide the $\KK{declassify}$ instruction, which
transfers a $\K{secret}$ value to its equivalent $\K{public}$ one.
Both $\KK{classify}$ and $\KK{declassify}$ exist purely to make explicit any
changes in security status; as Fig.~\ref{fig:reduction} shows, these
instructions do not imply any runtime cost.
These security casting operations (and our annotations, in general) do,
however, slightly increase the size of the bytecode when dealing with
$\K{secret}$ values (purely $\K{public}$ computations are unaffected), but the
simplicity and explicit nature of the security annotations are a worthwhile
trade-off.
We give experimental bytecode results for our \ctwasm cryptographic
implementations in Section~\ref{sec:implementation:eval}.

To restrict the use of $\KK{declassify}$, as Fig.~\ref{fig:typing} shows, we
extend function types with a \textit{trust} annotation that specifies whether
or not the function is $\K{trusted}$ or $\K{untrusted}$.
In turn, \ctwasm ensures that only $\K{trusted}$ functions may use
$\KK{declassify}$ and escape the restrictions (and guarantees) of the \ctwasm
type system.
For $\K{untrusted}$ functions, any occurrence of $\KK{declassify}$ is an error.
Moreover, trust is transitive: an $\K{untrusted}$ function is not permitted to
call a $\K{trusted}$ function.

We enforce these restrictions in the typing rules for $\KK{call}$ and
$\KK{call\_indirect}$.
But, per the original WebAssembly specification, the \KK{call\_indirect}
instruction must be additionally guarded by a runtime type check to ensure type
safety.
Thus we extend this runtime type check to additionally check that security
annotations are respected.
This is the only place in the semantics where our security annotations have any
effect on runtime behavior.

Put together, our security restrictions allow \ctwasm to communicate strong
guarantees through its types.
In an $\K{untrusted}$ function, where $\KK{declassify}$ is disallowed, it is
impossible for a $\K{secret}$ value to be directly or indirectly used in a way
that can reveal its value.
Thus, sensitive information such as private keys can be passed into unknown,
web-delivered functions, and so long as the function can be validated as
$\K{untrusted}$, \ctwasm guarantees that it will not be leaked by the action of
the function.
We next describe our mechanization effort, which includes a proof of this
property (see Section~\ref{typeconstanttime}).

\section{Formal Model}
\label{sec:model}

We provide a fully mechanized model of the \ctwasm language, together with several mechanized proofs of important properties, including a full proof of soundness of the extended type system, together with proofs of several strong security properties relating to information flow and constant-time.
 We build on top of a previous Isabelle model of WebAssembly~\cite{Watt:2018:MVW:3176245.3167082}, extending it with typing rules incorporating our secret types, annotations for trusted and untrusted functions, and the semantics of classification and declassification.
 At a rough count, we inherit \textasciitilde{}8,600 lines of non-comment, non-whitespace Isabelle code from the existing mechanization, with our extensions to the semantics and soundness proofs representing \textasciitilde{}1,700 lines of alterations and insertions.
 Our new security proofs come to \textasciitilde{}4,100 lines.

\subsection{Soundness}
We extend the original mechanized soundness proof of the model to our enhanced type system.
 For the most part, this amounted to a fairly mechanical transformation of the existing proof script.
 While we re-prove both the standard \textit{preservation} and \textit{progress} soundness properties, we will not illustrate the progress property in detail here, since its proof remains almost unchanged from the existing work, while the preservation property is relevant to our subsequent security proofs, and required non-trivial changes for the cases relating to function calls.
Both proofs proceed by induction over the definition of the typing relation.

WebAssembly's top level type soundness properties are expressed using an extended typing rule given over configurations together with an instance, as a representation of the WebAssembly runtime state.
Broadly, a configuration $c = \textnormal{\x{s}; \x{vs}; \x{es}}$ is given a \textit{result type} of the form $\x{ts}$ if its operation stack, \x{es}, can be given a \textit{stack type} of the form $[] \rightarrow \x{ts}$ under a typing context $C$ which abstracts the instance, the store $\x{s}$, and local variables $\x{vs}$. This judgement is written as \mbox{$\ \vdash_i c \textnormal{ : \textit{ts}}$}.

 We further extend this so that configurations, formerly typed by \textit{ts}, the \textit{result type} of their stack, are additionally typed according to the level of trust required for their execution; configuration types now take the form (\textit{tr, ts}).
For example, a configuration containing the privileged $\KK{declassify}$ operation will have ``$\K{trusted}$'' as the trust component of its type.
 The preservation property now certifies that trust is preserved by reduction along with the type of the configuration's stack.
 As a consequence, a configuration that is initially typed as $\K{untrusted}$ is proven to remain typeable as $\K{untrusted}$ across its entire execution, and will never introduce a privileged instruction at any intermediate stage of reduction.

\begin{theorem1}[preservation]
Given a configuration c, if \mbox{$\ \vdash_i c \textnormal{ : (\textit{tr, ts})}$} and \mbox{ $c \overset{a}{\leadsto}_i c'$}, then \mbox{$\vdash_i c' \textnormal{ : (\textit{tr, ts})}$}.
\end{theorem1}

\subsection{Security Properties}
\label{sec:security}
We provide fully mechanized proofs, in Isabelle, that our type system guarantees several related language-level security properties for all $\K{untrusted}$ code.
These proofs, as well as the full definition of the leakage model, are available in~\cite{supplement}.
We show that \ctwasm's type system guarantees several security properties, including non-interference and constant-time.
We conclude by showing that a well-typed $\K{untrusted}$ \ctwasm program is guaranteed to satisfy our constant-time property, the property which was the motivation for the type system's design.

Provided definitions, lemmas, and theorems are directly named according to their appearances in the mechanization, for easy reference.

\subsection{Public Indistinguishability}

\begin{figure*}
\footnotesize\fontsize{9pt}{9pt}


  $t_1\KK{.const}~k_1 \simval t_2\KK{.const}~k_2 \triangleq t_1 = t_2 \wedge (k_1 = k_2 \vee \K{sec}~t_1 =\K{sec}~t_2 = \K{secret})$

$$
\begin{array}{@{}r@{~}c@{~}l@{}}
  e_1 \simexp e_2 & \triangleq &
    \begin{cases}
\vspace{0.3cm}
      (\x{e_a} \simval \x{e_b}) & \text{if\hspace{0.5cm}} \pbox{0.23\textwidth}{\text{$e_1 = t_1\KK{.const}~k_1$} \linebreak \pbox{0.25\textwidth}{\text{$e_2 = t_2\KK{.const}~k_2$}}} \\

\vspace{0.3cm}
      (\x{e_a} \simexp \x{e_b})^n & \text{if\hspace{0.5cm}} \pbox{0.23\textwidth}{\text{$e_1 = \KK{block}~\x{ft}~\x{e_a}^n~\KK{end}$} \linebreak \pbox{0.25\textwidth}{\text{$e_2 = \KK{block}~\x{ft}~\x{e_b}^n~\KK{end}$}}} \text{or\hspace{0.4cm}} \pbox{0.25\textwidth}{\text{$e_1 = \KK{loop}~\x{ft}~\x{e_a}^n~\KK{end}$} \linebreak \pbox{0.2\textwidth}{\text{$e_2 = \KK{loop}~\x{ft}~\x{e_b}^n~\KK{end}$}}}\\

\vspace{0.3cm}
\pbox{0.25\textwidth}{
      $(\x{e_a} \simexp \x{e_b})^n$ \\
\pbox{0.25\textwidth}{
      $\wedge~(\x{e_c} \simexp \x{e_d})^m$
}
}
& \text{if\hspace{0.5cm}} \pbox{0.27\textwidth}{\text{$e_1 = \KK{if}~\x{ft}~\x{e_a}^n~\KK{else}~\x{e_c}^m~\KK{end}$} \linebreak \pbox{0.25\textwidth}{\text{$e_2 = \KK{if}~\x{ft}~\x{e_b}^n~\KK{else}~\x{e_d}^m~\KK{end}$}}} \text{or\hspace{0.4cm}} \pbox{0.25\textwidth}{\text{$e_1 = \KK{label}_n\{\x{e_a}^n\}~\x{e_c}^m~\KK{end}$} \linebreak \pbox{0.2\textwidth}{\text{$e_2 = \KK{label}_n\{\x{e_b}^n\}~\x{e_d}^m~\KK{end}$}}}\\

\vspace{0.3cm}
\pbox{0.25\textwidth}{
      $(\x{v_a} \simval \x{v_b})^n$ \\
\pbox{0.25\textwidth}{
      $\wedge~(\x{e_a} \simexp \x{e_b})^m$
}
}
& \text{if\hspace{0.5cm}} \pbox{0.25\textwidth}{\text{$e_1 = \KK{local}_n\{i;\x{v_a}^n\}~\x{e_a}^m~\KK{end}$} \linebreak \pbox{0.25\textwidth}{\text{$e_2 = \KK{local}_n\{i;\x{v_b}^n\}~\x{e_b}^m~\KK{end}$}}}\\

      e_1 = e_2 & \text{otherwise}
    \end{cases}
\end{array}
$$

$$
\begin{array}{@{}r@{\hspace{0.2cm}}c@{\hspace{0.2cm}}l@{}}
\left\{
\begin{array}{@{}l@{}}
\x{inst_1}^\ast,\\
\x{func_1}^\ast,\\
(\x{mut_1}, \x{glob_1})^\ast,\\
\x{table_1}^?,\\
(\x{sec_1}, \x{mem_1})^?
\end{array}
\right\}
~\simstr
\left\{
\begin{array}{@{}l@{}}
\x{inst_2}^\ast,\\
\x{func_2}^\ast,\\
(\x{mut_2}, \x{glob_2})^\ast,\\
\x{table_2}^?,\\
(\x{sec_2}, \x{mem_2})^?
\end{array}
\right\}
& \triangleq &
\begin{array}{@{}l@{}}
(\x{inst_1} = \x{inst_2})^\ast\\
\wedge~(\x{func_1} = \x{func_2})^\ast\\
\wedge~(\x{mut_1} = \x{mut_2} \wedge \x{glob_1} \simval \x{glob_2})^\ast\\
\wedge~(\x{table_1} = \x{table_2})^?\\
\wedge~\left(
\begin{array}{l}
(\K{size}~\x{mem_1} = \K{size}~\x{mem_2} \wedge \x{sec_1} = \x{sec_2} = \K{secret}) \\
\vee~ (\x{mem_1} = \x{mem_2} \wedge \x{sec_1} = \x{sec_2} = \K{public})
\end{array}
\right)^?
\end{array}
\end{array}
$$

$  \x{s_1}; \x{v_1}^\ast; \x{e_1}^\ast~\simcfg~\x{s_2}; \x{v_2}^\ast; \x{e_2}^\ast \triangleq (\x{s_1} \simstr \x{s_2}) \wedge (\x{v_1} \simval \x{v_2})^\ast \wedge (\x{e_1} \simexp \x{e_2})^\ast$

\caption{Definition of $\simcfg$.}
\vspace{-1\baselineskip}
\label{fig:config-equiv}

\end{figure*}

We define an indistinguishability relation between WebAssembly configurations, given by $\simcfg$.
 Intuitively, \textit{(public) indistinguishability} holds between two configurations if they differ only in the values of their secret state.
 That is, the values and types of their public state must be equal, as must the types of their secret state.
 Formally, we define $\simcfg$ over configurations in terms of indistinguishability relations for each of their components.
 These definitions can be found in Fig.~\ref{fig:config-equiv}.
 This relation is required for the expression of the constant-time property, and mirrors the equivalence relation used for the same purpose by~\cite{Barthe:2014:SNC:2660267.2660283} between program states.
  We prove that typeability of a WebAssembly configuration is invariant with respect to $\simcfg$.

\begin{lemma1}[equivp\_config\_indistinguishable]
$\simcfg$ is an equivalence relation.\end{lemma1}

\begin{lemma1}[config\_indistinguishable\_imp\_config\_typing]\label{lemma:config_typing}
If \mbox{$\ \vdash_i \textit{c} \textnormal{ : (\textit{tr, ts})}$}, then for all $c'$ such that $c \simcfg c'$, \mbox{$\ \vdash_i \textit{$c'$} \textnormal{ : (\textit{tr, ts})}$}.
\end{lemma1}

\subsection{Action Indistinguishability}

\begin{figure*}
\fontsize{9pt}{9pt}
$$
\begin{array}{@{}rcl@{}}
s; \x{vs};~(\K{s32}.\KK{const}~k_1)~(\K{s32}.\KK{const}~k_2)~\K{s32.}\x{binop}&\overset{a}{\leadsto}_i&
  s; \x{vs};~(\K{s32}.\KK{const}~(k_1~\x{binop}~k_2))
\end{array}
$$
\vspace{-1.3\baselineskip}
$$
\begin{array}{@{}rcl@{}}
\text{with } a & \triangleq & \K{binop\_action}(\x{binop}, (\K{s32}.\KK{const}~k_1), (\K{s32}.\KK{const}~k_2))
\end{array}
$$
\vspace{-0.5\baselineskip}
$$
\begin{array}{@{}rclr@{}}
s; \x{vs};~\x{v}_1^n~\KK{callcl}~\x{cl}&\overset{a}{\leadsto}_i&
  s'; \x{vs}';~\x{v}_2^m & (*)
\end{array}
$$
\vspace{-0.8\baselineskip}
$$
\begin{array}{@{}rcl@{}}
\text{with } \x{cl} & \triangleq & \{{\K{type}~(\x{tr}, t_1^n \to t_2^m), \K{host}}~...\}\\
a & \triangleq & \K{host\_action}(\x{s}, \x{v}_1^n, \x{s}', \x{v}_1^m, \x{cl})
\end{array}
$$

$$
\begin{array}{@{}r@{~}c@{~}l@{}}
  a_1 \simact a_2 & \triangleq &
    \begin{cases}
\vspace{0.2cm}
      \x{op}_1 = \x{op}_2 & \text{if\hspace{0.5cm}} \pbox{0.32\textwidth}{\text{$a_1 = \K{binop\_action}(\x{op}_1, \x{v}_1, \x{v}_2$)} \linebreak \pbox{0.25\textwidth}{\text{$a_2 = \K{binop\_action}(\x{op}_2, \x{v}_1', \x{v}_2'$)}}} \text{and\hspace{0.3cm}is\_safe\_binop($\x{op}_1$)}\\


\vspace{0.1cm}
\pbox{0.25\textwidth}{
      $\x{s}_1 \simstr \x{s}_2$ \\
\pbox{0.25\textwidth}{
      $\wedge~(\x{v}_1 \simval \x{v}_2)^n$
}\\
\pbox{0.25\textwidth}{
      $\wedge~\x{s}_1' \simstr \x{s}_2'$
}\\
\pbox{0.25\textwidth}{
      $\wedge~(\x{v}_{1'} \simval \x{v}_{2'})^m$
}\\
\pbox{0.25\textwidth}{
      $\wedge~\x{cl}_1 = \x{cl}_2$
}
}
& \text{if\hspace{0.5cm}} \pbox{0.35\textwidth}{\text{$a_1 = \K{host\_action}(\x{s}_1, \x{v}_1^n, \x{s}', \x{v}_{1'}^m, \x{cl}_1)$} \linebreak \pbox{0.25\textwidth}{\text{$a_2 = \K{host\_action}(\x{s}_2, \x{v}_2^n, \x{s}_2', \x{v}_{2'}^m, \x{cl}_2)$}}} \text{and\hspace{0.3cm}trust($\x{cl}_1$) = \K{untrusted}}\\
\vspace{0.1cm}
... \\
\vspace{0.2cm}
      a_1 = a_2 & \text{otherwise}
    \end{cases}
\end{array}
$$
\vspace{-0.2\baselineskip}
$$
\begin{array}{@{}ll@{}}
(*) & \pbox{0.75\textwidth}{\textnormal{The full axiomatic description of host function behavior is not reproduced here. Full details can be found in~\cite{Watt:2018:MVW:3176245.3167082}, or in our mechanization.}}
\end{array}
$$
\vspace{-0.8\baselineskip}
\caption{Example of \ctwasm action annotations and equivalence.}
\vspace{-1\baselineskip}
\label{fig:action}

\end{figure*}

The constant-time property is most naturally expressed as an equivalence of \textit{observations}, which are abstractions of an attacker's knowledge defined with respect to the \textit{leakage model} of the system.
We adopt a leakage model which extends the leakiest model depicted by~\cite{ctverif}, accounting for leakage of branch conditions, memory access patterns, and the operand sizes of unsafe binary operations, namely division and modulus.
 In addition, we must express our trust in the host environment that WebAssembly is embedded within.
 A host function marked as $\K{untrusted}$ will leak all public state it has access to when called, but never any secret state.
 In reality, the host environment is the web browser's \js engine, and user-defined \js is treated as \K{trusted}, so this corresponds to trusting that the engine's provided built-in functions are not malicious or compromised, and obey the properties guaranteed by the $\K{untrusted}$ annotation.

We augment the WebAssembly reduction relation with state-parameterized actions, as \mbox{$c \overset{a}{\leadsto}_i c'$}, effectively defining a labelled transition system.
Traditionally, a constant-time proof in the style of~\cite{Barthe:2014:SNC:2660267.2660283} would define its leakage model as a function from either action or state to a set of observations.
However, it is instead convenient for us to adopt a novel representation of the leakage model as an equivalence relation, given by $\simact$, between actions, denoting \textit{action indistinguishability}.
Intuitively, if two actions are defined as being equivalent by $\simact$, this implies that they are indistinguishable to an attacker.
This definition is inspired by the \textit{low view} equivalence relations seen in formal treatments of information flow~\cite{Sabelfeld:2006:LIS}, which are used to embody an attacker's view of a system.
An illustration of our definitions can be found in Fig.~\ref{fig:action}.

This approach is helpful because the behavior of the \ctwasm host environment, as inherited from \wasm, is specified entirely axiomatically, and may leak a wide variety of differently-typed state, making a set-based definition of leakage unwieldy.
For completeness, we sketch a more traditional leakage model as a supplement in the mechanization, although this leakage model does not capture the full range of observations induced by the leakage of the host environment, because, as mentioned, such a definition would be overly complicated when we have a simpler alternative.

 Having chosen our equivalence-based representation of the leakage model, \textit{observations} become instances of a quotient type formed with respect to $\simact$.
 This notion will be made precise in Section \ref{quotient}.

\begin{lemma1}[equivp\_action\_indistinguishable]
$\simact$ is an equivalence relation.
\end{lemma1}

This representation allows us to define a configuration being constant-time as a property of trace equivalence with respect to $\simact$.
 However, one final issue must be ironed out.
 Taking informal definitions of ``trace'' and ``observation'' for illustrative purposes, the standard statement of the constant-time property for a WebAssembly configuration could na\"{i}vely read as follows:

\begin{sketchdefinition1}[na\"{i}ve constant-time]
A configuration-instance pair (c,i) is constant-time iff for all $c'$ such that $c \simcfg c'$, the trace of $(c,i)$ and the trace of $(c',i)$ induce the same observations.
\end{sketchdefinition1}

Unfortunately, WebAssembly is not a completely deterministic language, and so this standard definition does not apply, as a configuration cannot be uniquely associated with a trace.
 There are two ways we can address this.
First, we can alter the semantics of WebAssembly to make it deterministic.
But, despite WebAssembly's non-determinism being highly trivial in most respects, one of our goals is for \ctwasm to be a strict extension to WebAssembly's existing semantics.
 Instead, we choose to generalize the standard definition of constant-time so that it can be applied to non-deterministic programs, in an analogous way to known possibilistic generalizations of security properties such as non-interference~\cite{nondetthesis, Mantel:2000}.
  A formal statement and proofs related to this generalized definition will follow in Section \ref{typeconstanttime}.

\begin{sketchdefinition1}[non-deterministic constant-time]\label{nondetconstanttime}
A configuration-instance pair (c,i) is constant-time iff, for all $c'$ such that $c \simcfg c'$, the set of traces of $(c,i)$ and the set of traces of $(c',i)$ induce the same observations.
\end{sketchdefinition1}

This generalization implicitly introduces the assumption that, where more than one choice of reduction is available, the probability of a particular single step being chosen is not dependent on any secret state.
 For WebAssembly, we have very good reason to expect that this is the case, because, as previously mentioned, WebAssembly's non-determinism is highly trivial---also as a deliberate design decision.
 The only relevant non-determinism which exists in the model is the non-determinism of the \KK{grow\_memory} instruction, non-determinism of exception (\KK{trap}) propagation, and non-determinism of the host environment.
 For \KK{grow\_memory}, our type system forces all inputs and outputs of the operation to be \K{public}, and our leakage model specifies that the length of the memory is leaked by the operation.
 For exception propagation, WebAssembly's non-determinism in this aspect is purely an artifact of the formal specification's nature as a small-step semantics, and the definition of its evaluation contexts.
 In a real implementation, when an exception occurs, execution halts immediately.
 For the host, we simply trust that the user's web browser is correctly implemented and, when making non-deterministic choices, respects $\K{secret}$ and $\K{untrusted}$ annotations, with respect to our leakage model.

\subsection{Self-isomorphism}
We initially prove a security property for arbitrary untrusted sections of code which is a single-step analogy to the \textit{self-isomorphism} property~\cite{10.1007/978-3-642-35308-6_11}, which, stepwise comparing the executions of all program configurations with observably equivalent state, forbids observable differences not just in the state, but in the program counter.
This single-step property is very strong, and is the key to proving all of the future properties given in this section.
The proof proceeds by induction over the definition of the reduction relation.

\begin{lemma1}[config\_indistinguishable\_imp\_reduce]
If \mbox{$\ \vdash_i \textit{c} \textnormal{ : ($\K{untrusted}$\textit{, ts})}$} for some \textit{ts}, then for all $c'$ such that $c \simcfg c'$, if $c \overset{a}{\leadsto}_i c_a$ then there exists $c'_a$ and $a'$ such that $c' \overset{a'}{\leadsto}_i c'_a$ and $c_a \simcfg c'_a$ and $a \simact a'$.
\end{lemma1}

From the definition of $\simcfg$, we know that $c_a$ and $c'_a$ contain the same instructions, modulo the values of secretly typed constants.

\subsection{Bisimilarity}

We now define our notion of bisimilarity.
We prove that programs that vary only in their secret inputs are bisimilar to each-other while performing $\simact$-equivalent actions in lockstep.
This property is sometimes known as the \textit{strong security property}~\cite{Sabelfeld:2000:PNM}.
 Configurations in WebAssembly always reduce with respect to an instance, so we define bisimulation in terms of configurations together with their instances.

\begin{definition1}[config\_bisimulation]
$\textnormal{config\_bisimulation R} \triangleq \newline
\indent \forall ((c,i), (c',i')) \in \textnormal{R}.~\newline
\indent \indent (\forall c_a,~a.
~c \overset{a}{\leadsto}_i c_a \longrightarrow \exists c'_a,~a'.
~c' \overset{a'}{\leadsto}_{i'} c'_a \wedge a \simact a' \wedge (c_a,i), (c'_a,i') \in \textnormal{R}) \wedge  \newline
\indent \indent \indent (\forall c'_a,~a'.
~c' \overset{a'}{\leadsto}_{i'} c'_a \longrightarrow \exists c_a,~a.
~c \overset{a}{\leadsto}_i c_a \wedge a \simact a' \wedge (c_a,i), (c'_a,i') \in \textnormal{R})$
\end{definition1}

\begin{definition1}[config\_bisimilar]
$\textnormal{config\_bisimilar} \triangleq \bigcup~\{~\textnormal{R}~|~\textnormal{config\_bisimulation R}~\}$
\end{definition1}

We prove that the set of pairs of well-typed, publicly indistinguishable configurations forms a bisimulation.
 From this and our definition of bisimilarity, we immediately have our version of the strong security property.

\begin{definition1}[typed\_indistinguishable\_pairs]
$\textnormal{typed\_indistinguishable\_pairs} \triangleq \{~((c,i),(c',i))~|~\vdash_i \textit{c} \textnormal{ : ($\K{untrusted}$\textit{, ts})} \wedge c \simcfg c' ~\}$
\end{definition1}

\begin{lemma1}[config\_bisimulation\_typed\_indistinguishable\_pairs]
$\textnormal{config\_bisimulation typed\_indistinguishable\_pairs}$
\end{lemma1}

\begin{theorem1}[config\_indistinguishable\_imp\_config\_bisimilar]
If \mbox{$\ \vdash_i \textit{c} \textnormal{ : ($\K{untrusted}$\textit{, ts})}$} for some \textit{ts}, then for all $c'$ such that $c \simcfg c'$, $((c,i), (c',i)) \in \textnormal{config\_bisimilar}$..
\end{theorem1}

\subsection{Non-interference}

We define a reflexive, transitive version of our reduction relation, given as $c \overset{as}{\leadsto}^*_i c_{as}$, annotated by an ordered list of actions.
We can then prove the following property as a transitive generalization of our initial lemma, capturing the classic non-interference property.
This is a strict information flow input-output property which encodes that publicly indistinguishable programs must have publicly indistinguishable outputs.

\begin{lemma1}[config\_indistinguishable\_trace\_noninterference]
If \mbox{$\ \vdash_i \textit{c} \textnormal{ : ($\K{untrusted}$\textit{, ts})}$} for some \textit{ts}, then for all $c'$ such that $c \simcfg c'$, if $c \overset{as}{\leadsto}^*_i c_{as}$ then there exists $c'_{as}$ and $as'$ such that $c' \overset{as'}{\leadsto}^*_i c'_{as}$ and $c_{as} \simcfg c'_{as}$ and $as$ pairwise $\simact$ with $as'$.
\end{lemma1}

\subsection{Constant-time}\label{typeconstanttime}

We now formally discuss the constant-time property we originally sketched (Section \ref{nondetconstanttime}).
We define, coinductively, the set of possible traces for a configuration with respect to an instance.
In Isabelle, the trace is represented by the type \textit{action llist}, the codatatype for a potentially infinite list of actions.
Equivalence between traces is then given as corecursive pairwise comparison by $\simact$, written as llist\_all2 $\simact$ in Isabelle.
We lift equivalence between traces to equivalence between sets of traces in the standard way.
This is already defined as a specialization of Isabelle's built-in rel\_set predicate. 

\begin{definition1}[config\_is\_trace]
$ \begin{array}{@{\hspace{0mm}}r@{\;}l@{\hspace{0mm}}} 
\nexists c_a.~c \overset{a}{\leadsto}_i c_a & \longrightarrow \textnormal{config\_is\_trace}~(c,i)~\textnormal{[]}\\
c \overset{a}{\leadsto}_i c_a \wedge \textnormal{config\_is\_trace}~(c_a,i)~as & \longrightarrow \textnormal{config\_is\_trace}~(c,i)~(a::as)
    \end{array} 
$ \hfill \phantom{.} 
\end{definition1}

\begin{definition1}[config\_trace\_set]
$\textnormal{config\_trace\_set}~(c,i) \triangleq \{~as~|~\textnormal{config\_is\_trace}~(c,i)~as~\}$
\end{definition1}

\begin{definition1}[rel\_set]
$\textnormal{rel\_set } R~A~B \triangleq (\forall x \in A.~ \exists y \in B.~ R~x~y) \wedge (\forall y \in B.~ \exists x \in A.~ R~x~y)$
\end{definition1}

\begin{definition1}[trace\_set\_equiv]
$\textnormal{trace\_set\_equiv} \triangleq \textnormal{rel\_set (llist\_all2 $\simact$)}$
\end{definition1}

From the above, we can now formally define our constant-time property.
We establish \ctwasm's titular theorem: all typed $\K{untrusted}$ configurations are constant-time.

\begin{definition1}[constant\_time\_traces]
$\textnormal{constant\_time\_traces}~(c,i) \triangleq \newline
\indent \forall c'.
~c \simcfg c' \longrightarrow \textnormal{trace\_set\_equiv } (\textnormal{config\_trace\_set }(c,i))~(\textnormal{config\_trace\_set } (c',i))$
\end{definition1}

\begin{theorem1}[config\_untrusted\_constant\_time\_traces]
If \mbox{$\ \vdash_i \textit{c} \textnormal{ : ($\K{untrusted}$\textit{, ts})}$} for some \textit{ts}, then (c,i) is constant-time.
\end{theorem1}

\subsection{Observations as Quotient Types}\label{quotient}
The definition above gives the constant-time property in terms of an equivalence between trace sets, where the abstract observations of existing literature on the constant-time property are left implicit in the definition of $\simact$.
We now discuss how observations can be re-introduced as objects into our formalism, allowing us to adopt the standard definition of the constant-time property as equality between sets of observations.

We observe that, as $\simact$ is an equivalence relation, we may use it to define a quotient type~\cite{quotienttypes}.
Quotient types are the type-theoretic analogy to quotient sets, where elements are partitioned into equivalence classes.
 Isabelle allows us to define and reason about quotient types, and to verify that particular functions over the underlying type may be \textit{lifted} to the quotient type and remain well-defined~\cite{Huffman2013LiftingAT}.
This amounts to a proof that the function has the same value for each member of an equivalence class abstracted by the quotient type.

We can define the type of \textit{observations} as the quotient type formed from the underlying type \textit{action llist} with the equivalence relation being llist\_all2 $\simact$.
 Since $\simact$ defines our leakage model, this \textit{observation} type precisely characterizes the information that the model allows an attacker to observe during execution.
We can then (trivially) lift the previous configuration trace set definition to observations, and give our alternative definition of the constant-time property.

\begin{definition1}[observation]\label{trace}
$\textit{observation} \triangleq \textit{action llist}~/~(\textnormal{llist\_all2}~\simact)$
\end{definition1}

\begin{lemma1}[config\_obs\_set]
The lifting of the function \textit{config\_trace\_set} from the type \mbox{((config $\times$ inst) $\rightarrow$ (action llist) set)} to the type \mbox{((config $\times$ inst) $\rightarrow$ observation set)} is well-defined.
\end{lemma1}

\begin{definition1}[constant\_time]
$\textnormal{constant\_time}~(c,i) \triangleq \forall c'.
~c \simcfg c' \longrightarrow \textnormal{config\_obs\_set}~(c,i) = \textnormal{config\_obs\_set}~(c',i)$
\end{definition1}

We additionally give weaker versions of all of the results in \ref{typeconstanttime} and \ref{quotient} using a stronger definition of config\_is\_trace that is inductive rather than coinductive in~\cite{supplement}.
\section{Implementation}
\label{sec:implementation}
In this section we describe our \ctwasm implementations and supporting tools.
We also describe our evaluation of the \ctwasm language design and
implementation, using several cryptographic algorithms as case studies.
All materials referenced here are available in~\cite{supplement}.

\subsection{\ctwasm Implementations}
\label{sec:implementation:native}

We provide two \ctwasm implementations: a reference implementation and a native
implementation for V8 as used in both Node.js and the Chromium browser.
We describe these below.

\subsubsection{Reference implementation}
\label{sec:implementation:native:ref}
We extend the \wasm reference interpreter~\cite{wasm_ref_interpreter} to implement the
full \ctwasm semantics.
Beyond providing an easily understandable implementation of the spec, the
reference interpreter serves two roles.
First, it provides an easy to understand implementation of the \ctwasm
specification in a high-level language (OCaml), when compared to, say, the
optimized V8 implementation.
Moreover, the interpreter (unlike V8) operates on both bytecode and text-format
\ctwasm code.
We found this especially useful for testing handwritten \ctwasm crypto
implementations and our V8 implementation of \ctwasm.
Second, the reference \wasm implementation also serves as the basis for a
series of tools.
In particular, we reuse the parsers, typed data structures, and testing
infrastructure (among other parts) to build and test our supporting tools
and verified type checker.

\subsubsection{V8 implementation}
\label{sec:implementation:native:v8}
WebAssembly in both Node.js and Chromium is implemented in the V8 \js engine.
V8 parses \wasm bytecode, validates it, and directly compiles the bytecode to a
low-level ``Sea of Nodes''~\cite{Click:1995:SGI:202529.202534} representation (also used by the \js
just-in-time compiler), which is then compiled to native code.
We extend V8 (version 6.5.254.40) to add support for \ctwasm.
We modify the \wasm front-end to parse our extended bytecode and
validate our new types.
We modify the back-end to generate code for our new instructions.
While the parser modifications are straightforward, our validator fundamentally
changes the representation of types.
V8 assumes a one-to-one correspondence between the (Sea of Nodes) machine
representation of types and \wasm types.
This allows V8 to simply use type aliases instead of tracking \wasm types
separately.
Since $\K{s32}$ and $\K{s64}$ have the same machine representation as $\K{i32}$
and $\K{i64}$, respectively, our implementation cannot do this.
Our \ctwasm implementation, instead, tracks types explicitly and converts
\ctwasm types to their machine representation when generating code;
since our approach is largely type-driven, the code generation for \ctwasm is
otherwise identical to that of \wasm.
By inspecting the generated assembly code, we observed that V8 does not
compile the $\KK{select}$ instruction to constant-time assembly. We therefore
implement a separate instruction selection for $\K{secret}$ $\KK{select}$ so that the generated
code is in constant-time.

\ctwasm represents each instruction over secrets as a two-byte
sequence---the first byte indicates if the operation is over a secret, the
second indicates the actual instruction.
We take this approach because the existing, single-byte instruction space is
not large enough to account for all (public and secret) \ctwasm instructions;
introducing polymorphism is overly intrusive to the specification and
V8 implementation.
Importantly, this representation is backwards compatible: all public operations
are encoded in a single byte, as per the \wasm spec.
Indeed, all our modifications to the V8 engine preserve backwards
compatibility---\ctwasm is a strict superset of \wasm, and thus our changes do
not affect the parsing, validation or code generation of legacy \wasm code.

\subsection{Verified Type Checker}
\label{sec:implementation:validator}

We provide a formally verified type checker for \ctwasm stacks, and integrate it with our extension of the OCaml reference implementation.
 This type checker does not provide the informative error messages of its unverified equivalent, so we include it as an optional command-line switch which toggles its use during the validation phase of \ctwasm execution.
We validate this type checker against our conformance tests, and our crypto implementations.

The type checker is extended from the original given by~\cite{Watt:2018:MVW:3176245.3167082}, however major modifications needed to be made to the original constraint system and proofs.
The original type checker introduced an enhanced type system with polymorphic symbols; the type of an element of the stack during type checking could either be entirely unconstrained (polymorphic), or an exact value type.
We must add an additional case to the constraint system in order to produce a sound and complete algorithm; it is possible for an element of the stack to have a type that is unconstrained in its representation, but must be guaranteed to be secret.
This means that in addition to the original $\K{TAny}$ and $\K{TSome}$ constraint types, we must introduce the additional $\K{TSecret}$ type, and extend all previous lemmas, and the soundness and completeness proofs, for this case.

\subsection{\ctwasm Developer Tools}
\label{sec:implementation:tools}
We provide two tools that make it easier for developers to use \ctwasm:
\strip, allows developers to use \ctwasm as a development language
that compiles to existing, legacy \wasm runtimes;
\inferer, on the other hand, helps developers rewrite existing \wasm code to
\ctwasm.
We describe these below.

\subsubsection{\strip}
\label{sec:implementation:tools:strip}

Constant-Time WebAssembly is carefully designed to not only enforce security
guarantees purely by the static restrictions of the type system, but also be a
strict syntactic and semantic superset of WebAssembly.
These facts together mean that \ctwasm can be used as a principled development
language for cryptographic algorithms, with the final implementation
distributed as base WebAssembly with the crypto-specific annotations removed.
In this use-case, \ctwasm functions as a security-oriented analogy to
TypeScript~\cite{typescript}. TypeScript is a form of statically typed
JavaScript, designed to facilitate a work-flow where a developer can complete
their implementation work while enjoying the benefits of the type system,
before \textit{transpiling} the annotated code to base JavaScript for
distribution to end users.
Similarly, \ctwasm facilitates a work-flow where cryptography implementers can
locally implement their algorithms in Constant-Time WebAssembly in order to take
advantage of the information flow checks and guarantees built into our type
system, before distributing the final module as base WebAssembly.

We implement a tool, \strip, analogous to the TypeScript compiler,
for transpiling \ctwasm code to bare \wasm.
This tool first runs the \ctwasm type checking algorithm, then strips security
annotations from the code and removes the explicit coercions between
$\K{secret}$ and $\K{public}$ values.
Moreover, all $\K{secret}$ $\KK{select}$ operations are rewritten to an equivalent
constant-time sequence of bitwise operations, since, as previously mentioned in
Section \ref{sec:implementation:native:v8}, the $\KK{select}$ instruction is not always
compiled as constant-time.
Like the TypeScript compiler~\cite{typescriptunsound}, \strip does \emph{not}
guarantee the total preservation of all \ctwasm semantics and language
properties after translation, especially in the presence of other bare \wasm
code not originally generated and type checked by our tool.
However, we can offer some qualified guarantees even after translation.
 
With the exception of the \KK{call\_indirect} instruction, the runtime
behaviors of \ctwasm instructions are not affected by their security
annotations, as these are used only by the type system.
The \KK{call\_indirect} instruction exists to facilitate a dynamic function
dispatch system emulating the behavior of higher order code, a pattern
which is, to the best of our knowledge, non-existent in serious cryptographic
implementations.
Aside from this, bare WebAssembly interfacing with the generated code may
violate some assumptions of the Constant-Time WebAssembly type system.
For example, if the original code imports an $\K{untrusted}$ function, it assumes
that any $\K{secret}$ parameters to that function will not be leaked.
However at link-time, the type-erased code could have its import satisfied by a
bare WebAssembly function which does not respect the $\K{untrusted}$ contract.

\strip detects these situations, and warns the developer wherever the
translation may not be entirely semantics-preserving.
We aim for a sound overapproximation, so that a lack of warnings can give
confidence to the implementer that the translation was robust, but nevertheless
one that is realistic enough that many \ctwasm cryptographic implementations
can be transpiled without warnings (see Section~\ref{sec:implementation:eval}).

By default, \strip assumes that the host itself is a $\K{trusted}$ environment.
This matches the assumptions made throughout the paper.
The tool offers an additional \emph{paranoid} mode, which warns the
developer about every way the module falls short of total encapsulation.
These conditions are likely to be too strict for many real-world cryptographic
implementations designed to be used in a \js environment---the conditions imply
that the host is not allowed direct access to the buffer where the encrypted
message is stored.
But, as \wasm becomes more ubiquitous, this mode could provide additional
guarantees to self-contained \wasm applications (e.g., the Nebulet
micro-kernel~\cite{nebulet}) that do not rely on a \js host to execute.


\subsubsection{\inferer}
\label{sec:implementation:tools:inferer}

\ctwasm is a useful low-level language for implementing cryptographic
algorithms from the start, much like qhasm~\cite{qhasm} and
Jasmin~\cite{jasmin}.
But, unlike qhasm and Jasmin, WebAssembly is not a domain-specific language and
developers may already have crypto \wasm implementations.
To make it easier for developers to port such \wasm implementations to \ctwasm,
we provide a prototype tool, \inferer, that semi-automatically rewrites \wasm
code to \ctwasm.

At its core, \inferer implements an inference algorithm that determines the
security labels of local variables, functions, and globals.\footnote{
  \inferer operates at semantic level to allow non-local, cross-function
  inference, but also supports a syntactic mode which rewrites \wasm text
  format's S-expressions.
  We found both to be useful: the former in porting TEA and Salsa20 without
  manual intervention, the latter in semi-automatically porting the TweetNaCl
  library.
}
Our inference algorithm is conservative and initially assumes that every value is
$\K{secret}$.
It then iteratively traverses functions and, when assumptions are invalidated,
relabels values (and the operations on those values) to $\K{public}$ as
necessary.
For example, when encountering a $\KK{br\textunderscore{}if}$ instruction,
\inferer relabels the operand to $\K{public}$ and traverses the function AST
backwards to similarly relabel any of the values the operands it depends on.
For safety, our tool does not automatically insert any declassification instructions.
Instead, the developer must insert such instructions explicitly when
the label of a value cannot be unified.

Beyond manually inserting $\KK{declassify}$ instructions, \inferer also requires
developers to manually resolve the sensitivity of certain memory operations.
\inferer does not (yet) reason about memories that have mixed sensitivity data:
statically determining whether a memory load at a dynamic index is $\K{public}$
in the presence of $\K{secret}$ memory writes is difficult.
Hence, \inferer assumes that all memory is $\K{secret}$---it does not
create a separate module to automatically partition the $\K{public}$ and
$\K{secret}$ parts.
In such cases, the developer must resolve the type errors manually---a task we
found to be relatively easy given domain knowledge of the algorithm.
We leave the development of a more sophisticated tool (e.g., based on symbolic
execution~\cite{SurveySymExec-CSUR18}) that can precisely reason about memory---at least for
crypto implementations---to future work.

\subsection{Evaluation}
\label{sec:implementation:eval}

\newcolumntype{H}{S[round-mode=places, group-digits=false, table-format = 1.3,round-precision=3]}

\csvstyle{every csv}{
    respect all=true,
    late after line=\\,
    table foot={\bottomrule},
    command=\csvlinetotablerow
}

We evaluate the design and implementation of \ctwasm by answering the following
questions:
\begin{enumerate}
\item Can \ctwasm be used to express real-world crypto algorithms securely?
\item What is the overhead of \ctwasm?
\item Does \ctwasm (and \strip) produce code that runs in constant-time?
\end{enumerate}
\noindent To answer these questions, we manually implement three cryptographic
algorithms in \ctwasm: the Salsa20 stream cipher~\cite{salsa20}, the SHA-256
hash function~\cite{sha256}, and the TEA block cipher~\cite{tea}.\footnote{
  TEA and several variants of the block cipher are vulnerable and should not
  be used in practice~\cite{kelsey1997related, hong2003differential,
  hernandez2004finding}.
  We only implement TEA to evaluate our language as a measure of comparison
  with~\cite{Barthe:2014:SNC:2660267.2660283}.
}
Following~\cite{Barthe:2014:SNC:2660267.2660283}, we chose these three
algorithms because they are designed to be constant-time and
\emph{should} be directly expressible in \ctwasm.
Our implementations are straightforward ports of their corresponding C reference
implementations~\cite{c-salsa20, c-sha256, tea}.
For both Salsa20 and TEA, we label keys and messages as $\K{secret}$; for
SHA-256, like~\cite{Barthe:2014:SNC:2660267.2660283}, we treat the input
message as secret.
 
Beyond these manual implementations, we also port an existing \wasm
implementation of the TweetNaCl library~\cite{tweetnacl, tweetnacl-wasm}.
This library implements the full NaCl API~\cite{nacl}, which exposes 32 functions.
Internally, these functions are implemented using the XSalsa20, SHA-512,
Poly1305, and X25519 cryptographic primitives.
For this library, we use the \inferer tool to semi-automatically label values;
most inputs are $\K{secret}$, represented as ($\K{public}$) ``pointers'' into
the $\K{secret}$ memory.

We also use \strip to strip labels and produce fully unannotated versions all
our \ctwasm algorithms.
We run \strip with \emph{paranoid} mode off, since this corresponds to our
current security model.
\strip reported no warnings for any of the ports, i.e., no parts of the
translations were flagged as endangering the preservation of semantics, given
our previously stated assumptions.

To ensure that our ports are correct, we test our implementations against \js
counterparts.
For Salsa20 and SHA-256, we test our ports against existing \js
libraries, handling 4KB and 8KB inputs~\cite{npm-sha256, npm-salsa20}.
For TEA, we implement the algorithm in \js and test both our \ctwasm and \js
implementations against the C reference implementation, handling 8 byte
inputs~\cite{tea}.
Finally, for TweetNaCl, we use the \wasm library's test
suite~\cite{tweetnacl-wasm}.

\para{Experimental setup}
We run all our tests and benchmarks on a 24-core, 2.1GHz Intel Xeon 8160
machine with 1TB of RAM, running Arch Linux (kernel 4.16.13).
We use Node.js version 9.4.0 and Chromium version 65.0.3325.125, both using V8
version 6.5.254.40, for all measurements.
Unless otherwise noted, our reported measurements are for Node.js.
For each manually-ported crypto primitive we run the benchmark for 10,000
iterations, Salsa20 and SHA-256 processing 4KB and 8KB input messages, TEA
processing 8B blocks.
For TweetNaCl, we use the library's existing benchmarking infrastructure to run
each function for 100 iterations, since they process huge input messages
(approximately 23MB).
We report the median of these benchmarks.

\subsubsection{Expressiveness}
\label{sec:implementation:eval:expressiveness}
With \KK{declassify}, \ctwasm can trivially express any cryptographic algorithm,
even if inputs are annotated as \K{secret}, at the cost of security.
We thus evaluate the expressiveness of the \K{untrusted} subset of \ctwasm that
does not rely on \KK{declassify}.
In particular, we are interested in understanding to what degree real-world
crypto algorithms can be implemented as \K{untrusted} code and, when this is
not possible, if the use of \KK{declassify} is sparse and easy to audit.

We find that all crypto primitives---TEA, Salsa20, XSalsa20, SHA-256, SHA-512,
Poly1305, and X25519---can be implemented as \K{untrusted} code.
This is not very surprising since the algorithms are designed to be implemented
in constant-time.
Our port of Poly1305 did, however, require some refactoring to be fully
untrusted.
Specifically, we refactor an internal function (\texttt{poly1305\_blocks}) to
take a \K{public} value as a function argument instead of a reference to a
public memory cell (since our memory is \K{secret}).

The TweetNaCl library requires a single \KK{declassify} instruction, in
the \K{crypto\_secretbox} API---the API that implements secret-key
authenticated encryption.
As shown in Fig.~\ref{fig:declass}, we use $\KK{declassify}$ in the decryption
function (\K{crypto\_secretbox\_open}) to return early if the ciphertext fails
verification; this leak is benign, as the attacker already knows that any
modifications to the ciphertext will fail verification~\cite{nacl}.
A na\"{i}ve port of TweetNaCl (e.g., as automatically generated by \inferer) would
also require declassification in the \K{crypto\_sign\_open} API.
This function operates on public data---it performs public-key verification,
but relies on helper functions that are used by other APIs that compute on
secrets.
Since \ctwasm does not support polymorphism over secrets, the results of these
functions would need to be declassified.
Trading-off bytecode size for security, we instead refactor this API to a
separate \K{untrusted} module and copy these helper functions to compute on
public data.

\subsubsection{Overhead}
\label{sec:implementation:eval:overhead}
Using our TweetNaCl and our manually ported cryptographic algorithms, we
measure the overhead of \ctwasm on three dimensions---bytecode size, validation
time, and execution time.
Our extended instruction set imposes a modest overhead in bytecode size due to our new annotations
but imposes no overhead on non-cryptographic \wasm code.
We also find that \ctwasm does not meaningfully affect validation or runtime
performance.

\para{Bytecode size}
Since \ctwasm represents instructions over secrets as a two-byte
sequence, an annotated \ctwasm program will be as large or larger than its 
unannotated counterpart.
For the TweetNaCl library, the unannotated, original \wasm compiles to 21,662
bytes; the bytecode size of our semi-automatically annotated \ctwasm
version---including functions in the signing API that must be duplicated with
$\K{public}$ and $\K{secret}$ versions---is 40,050 bytes, an overhead of
roughly 85\%.
For our hand-annotated implementations of Salsa20, SHA256, and TEA, we measure the mean overhead to 
be 15\%.
The additional overhead for TweetNaCl is directly from the code
duplication---the overhead of an earlier implementation that used
\KK{declassify} was roughly 18\%---an overhead that can be reduced with
techniques such as label polymorphism~\cite{myers2001jif}.

\para{Validation}
\begin{figure}[t]
    \csvloop{
        file=eval/results/validation_timing.csv,
        before reading=\sisetup{table-alignment=center},
        tabular=lHHHHH,
        table head={%
            \toprule & \multicolumn{3}{c}{\ctwasm Node} & \multicolumn{2}{c}{Vanilla Node}\\
            \cmidrule(lr){2-4} \cmidrule(lr){5-6}
            & {\ctwasm} & {\wasm} & {\strip} & {\wasm} & {\strip}\\\midrule
        }
    }
  \vspace{0.5\baselineskip}
\caption{Median validation time (ms) of our ported crypto primitives and
TweetNaCl library. We report the performance of our \ctwasm port, the
original \wasm implementation, and the \strip stripped version for our
modified Node.js runtime and an unmodified, vanilla Node.}
\label{fig:eval-validation}
\end{figure}
We measure the performance of the \ctwasm type checker when validating both
annotated and unannotated (via \strip) code and compare its performance with an
unmodified validator.
Fig.~\ref{fig:eval-validation} summarizes our measurements.
We find that our baseline validator is 14\% slower than an unmodified
validator, a slowdown we attribute to our representation of \ctwasm types (see
Section~\ref{sec:implementation:native}).
Moving from unannotated code to annotated code incurs a cost of 20\%.
This is directly from the larger binary---validation in V8 is implemented as a
linear walk over the bytecode.
Note that though these relative slowdowns seem high, the absolute slowdowns
are sub-millisecond, only occur once in the lifetime of the program, and thus
have no meaningful impact on applications.

\pagebreak 

\para{Runtime}
\begin{figure}[t]
   \includegraphics{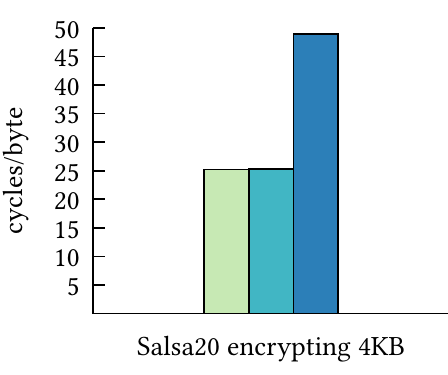}
   \includegraphics{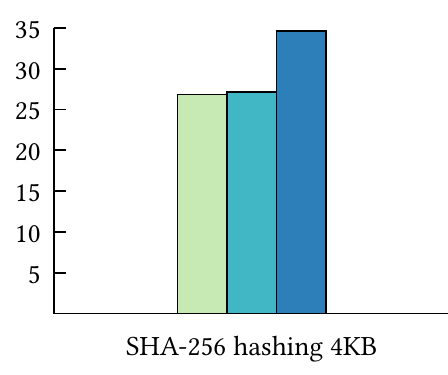}
   \includegraphics{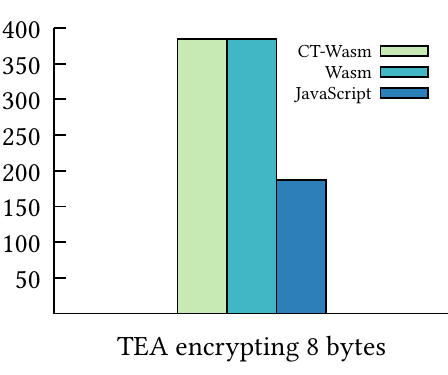}
  \vspace{-0.2\baselineskip}%
   \caption{Runtime performance of handwritten crypto primitives.}
   \label{fig:handperf}
  \vspace{-0.8\baselineskip}%
\end{figure}
As with validation, we measure the impact of both our modified engine and of
\ctwasm annotations on runtime performance.
We compare our results with reference \js implementations in the case of
Salsa20, SHA-256, and TEA; we compare our results with the reference TweetNaCl
\wasm implementation.
For both Node.js and Chromium we find that our modifications to the runtimes do
not impact performance and that \ctwasm generated code is on par with \wasm
code---the mean performance overhead is less than 1\%.
We find our TweetNaCl implementation to be as fast as the original
\wasm implementation for all the NaCl functions.
Fig.~\ref{fig:handperf} compares our manual ports with \js implementations:
\wasm and \ctwasm are comparable and faster than \js for both Salsa20 and
SHA-256.
The TEA \js implementation is faster than the \wasm counterparts; we believe
this is because the \js implementation can be more easily optimized
than \wasm code that requires crossing the \js-\wasm boundary.

\subsubsection{Security}
\label{sec:implementation:eval:security}
To empirically evaluate the security of our implementations, we run a modified
version of \dudect~\cite{Reparaz:2017:DMC:3130379.3130776}.
The \dudect tool runs a target program multiple times on different
input classes, collects timing information and performs statistical
analysis on the timing data to distinguish the distributions of the input
classes.
We modify \dudect to more easily use it within our existing \js infrastructure.
Specifically, we modify the tool to read timing information from a file and not
measure program execution times itself.
This allows us to record time stamps before and after running a crypto
algorithm, and ignore the effects of \js engine boot up time, \wasm validation,
etc.

\begin{figure}[hbt]
  \begin{minipage}[b]{0.49\linewidth}
    \centering
    \includegraphics[width=\linewidth]{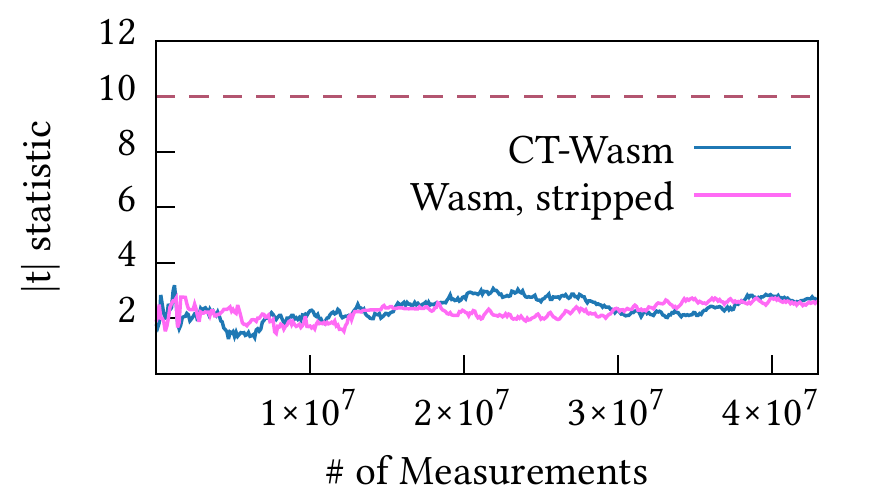}
    \subcaption{TweetNaCl secretbox implementation, \ctwasm and \strip}
    \label{fig:dudect-nacl-wasm}
  \end{minipage}\hspace{0.008\linewidth}
  \begin{minipage}[b]{0.49\linewidth}
    \centering
    \includegraphics[width=\linewidth]{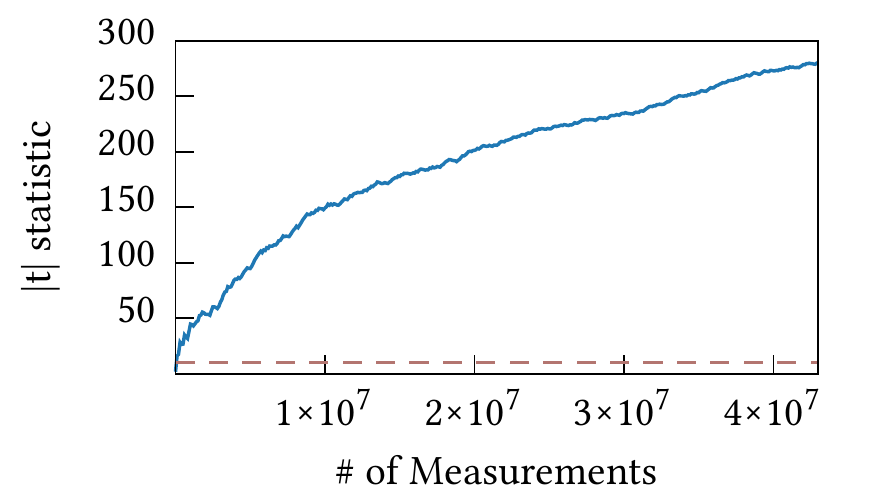}
    \subcaption{Salsa20 \ctwasm implementation, broken \js harness}
    \label{fig:dudect-salsa20-broken}
  \end{minipage}
  \caption{\texttt{dudect} measurements for various cryptographic algorithms.}
\vspace{-0.5\baselineskip}
  \label{fig:dudect-eval}
\end{figure}

We run our modified version of \dudect on the \ctwasm and \strip versions of
Salsa20, SHA-256, TEA, and TweetNaCl's secretbox API.
Following the methodology in~\cite{Reparaz:2017:DMC:3130379.3130776}, we
measure the timing of an all-zero key versus randomly generated keys (or
messages, in the case of SHA-256).
All other inputs are zeroed out.
We take 45 million measurements, each measurement running 10 iterations of the
respective algorithm.
For algorithms that can take an arbitrary message size, we use messages of 64
bytes in length.
\dudect compares the timing distributions of the two input
classes using the Welch's t-test, with a cutoff of $|t| > 10$ to disprove
the null hypothesis that the two distributions are the same.
As seen in Fig.~\ref{fig:dudect-nacl-wasm}, our \ctwasm and \strip
implementations for the TweetNaCl \textsf{secretbox} code have $|t|$ values
well below the threshold of 10; this is the case for all our other
algorithms as well.

Beyond ensuring that our \ctwasm implementations are constant-time, running
\dudect revealed the subtlety of using \js for crypto.
In an early implementation of the Salsa20 \js harness, we stored keys as arrays
of 32-bit integers instead of typed byte arrays before invoking the \ctwasm
algorithm.
As seen in Fig.~\ref{fig:dudect-salsa20-broken}, this version of the harness
was decidedly not constant-time.
We believe that time variability is due to \js transparently boxing/unboxing
larger integer values (e.g., those of the randomly generated keys), but
leaving smaller integer values alone (e.g., those of the all-zero key).

We also discovered a second interesting case while measuring the SHA-256 \js
implementation: calling the hash update function once per iteration, instead of
10, caused the timing distributions to diverge wildly, with $|t|$-statistics
well over 300.
Placing the function call inside a loop, even for just a single
iteration, caused the distributions to become aligned again, with $|t|$-statistics back under
the threshold of 10.
We did not observe this behavior for the \ctwasm SHA-256 implementation and hypothesize
that this time variability was due to \js function inlining.
We leave investigation of \js timing variabilities and their impact to future 
work.



\section{Related work}
\label{sec:related}
An initial high-level design for \ctwasm, which this work entirely supersedes,
has been previously described~\cite{priscctwasm}.

\para{Low-level crypto DSLs}
Bernstein's \textsf{qhasm}~\cite{qhasm} is an assembly-level language used to
implement many cryptographic routines, including the core algorithms of the
NaCl library. However, the burden is still on the developer to write constant-time
code, as \textsf{qhasm} has no notion of non-interference.
CAO~\cite{cao-lang} and Cryptol~\cite{cryptol} are higher-level DSLs for crypto implementations,
but do not have verified non-interference guarantees.

Vale~\cite{vale} and Jasmin~\cite{jasmin} are structured assembly languages
targeting high-performance cryptography, and have verification systems to prove
freedom from side-channels in addition to functional correctness. Vale and
Jasmin both target native machine assembly, and rely upon the Dafny
verification system~\cite{leino2010dafny}. Vale uses a flow-sensitive type system to enforce
non-interference, while Jasmin makes assertions over a constructed product
program with each compilation.
This work does not consider functional correctness in \ctwasm, and uses a very
simple type system to enforce non-interference. This approach scales
better in the context of a user's browser quickly verifying a downloaded
script for use in a web application.

\para{High-level crypto DSLs}
The HACL*~\cite{hacl} cryptographic library is written in constrained subsets
of the F* verification language that can be compiled to C.
Like \ctwasm, HACL* provides strong non-interference guarantees.
Unlike \ctwasm, though, the proof burden is on the developer and does not come
for free, i.e., it is not enforced by the type system directly.
Though it currently compiles to C, the HACL* authors are also targeting \wasm
as a compilation target~\cite{hacl-github}.
FaCT~\cite{fact} is a high-level language that compiles to LLVM which it then
verifies with \ctverif~\cite{ctverif}.
CAO~\cite{cao-lang, compiling-cao} and Cryptol~\cite{cryptol} are high-level DSLs for crypto implementations,
but do not have verified non-interference guarantees.
All these efforts are complementary to our low-level approach.

\para{Leakage models}
Our leakage model derives much of its legitimacy from
existing work on the side-channel characteristics of low-level languages, both
practical~\cite{Reparaz:2017:DMC:3130379.3130776,ctrules} and
theoretical~\cite{ctverif,Barthe:2014:SNC:2660267.2660283,10.1007/978-3-319-66402-6_16}.
We aim to express our top level security information flow and constant-time
properties in a way that is familiar to readers of these works.
Where our work differs from the above works on constant-time is in our
representation of \textit{observations}.
We draw inspiration from the equivalence relation-based formalizations
described by~\cite{Sabelfeld:2006:LIS} for timing sensitive non-interference, which
treats the number of semantic steps as its observation of a program execution.
This is fundamentally related, and sometimes even given as synonymous,
to the constant-time
condition~\cite{Barthe:2014:SNC:2660267.2660283}.

Our type system---which facilitates the non-interference result---can be characterized as
a specialization of the Volpano-Irvine-Smith security type system~\cite{Volpano:1996:STS:353629.353648}.
Our equivalence relation-based observations are similar to
the abstractions used by~\cite{cryptoeprint:2017:1233,barthe2018secure}.
To the best of our knowledge, our proof work is the first to use quotient types to connect such a low view
equivalence representation of an attacker's observational
power~\cite{Sabelfeld:2006:LIS} to a proof of a leakage model-based constant-time
property.

The literature on non-interference above is split as to whether traces and their
associated properties are expressed inductively or coinductively. We give
both interpretations, with the coinductive definition additionally capturing an
observation equivalence guarantee between publicly indistinguishable
non-terminating programs, encoding that even if a program does not terminate,
timing side-channels from visible intermediate side-effects will not leak
secret values.
~\cite{10.1007/978-3-642-35308-6_11} give a coinductive treatment of the
non-interference property, but for an idealized language, and do not connect it
to the constant-time property.


\section{Future Work}
\label{sec:future}
We have described two approaches to using \ctwasm, either as a native
implementation or a ``development language'' for base \wasm.
As an intermediate between these two, \ctwasm can be ``implemented'' in
existing engines by poly-filling the \texttt{WebAssembly} API to validate
\ctwasm code and rewrite it to \wasm.
Doing this efficiently is, unfortunately, not as simple as compiling \strip to
\js or WebAssembly---to avoid pauses due to validation, \strip must be
implemented efficiently (e.g., at the very least as a streaming validator).

\ctwasm takes a conservative approach to trust and secrecy polymorphism in order to ensure design
consistency with \wasm.
Even given this direction, there is possible space for relaxation,
especially regarding \KK{call\_indirect} and higher-order code.

While we have experimentally validated that our cryptography implementations do
not show input-dependent timing characteristics, the V8 WebAssembly
implementation is still relatively new.
Future implementations may implement aggressive optimizations that could
interfere with our guarantees.
%
A principled investigation of the possible implications of heuristically
triggered JIT optimizations on the timing characteristics of WebAssembly would
allow us to maintain our guarantees in the presence of more aggressive compiler
behaviours.

 
We foresee \ctwasm to be useful not only as a development language but also as
target language for higher-level crypto languages.
Since some of these language (e.g., HACL*~\cite{hacl} and FaCT~\cite{fact})
are already starting to target WebAssembly,
it would be fruitful extending these projects to target \ctwasm as a secure
target language instead.
At the same time, extending \inferer to (fully) automatically infer
security annotations from base \wasm would potentially prove yet more
useful---this would allow developers to compile C/C++ libraries such as
libsodium~\cite{libsodium} to \wasm (e.g., with Emscripten~\cite{emscripten-wasm}) and use
\inferer to ensure they are secure.


\section{Conclusion}
\label{sec:conclusion}
We have presented the design and implementation of Constant-Time WebAssembly, a
low-level bytecode language that extends WebAssembly to allow developers to
implement verifiably secure crypto algorithms.
\ctwasm is fast, flexible enough to implement real-world crypto libraries,
and both mechanically verified and experimentally measured to
produce constant-time code.
Inspired by TypeScript, \ctwasm is designed to be usable today, as a
development language for existing, base \wasm environments.
Both as a native and development language, \ctwasm provides a principled
direction for improving the quality and auditability of web platform
cryptography libraries while maintaining the convenience that has made \js
successful.

\begin{acks}                            
We thank the anonymous POPL and POPL AEC reviewers for their suggestions and
insightful comments.
We thank Andreas Rossberg and Peter Sewell for their support during this work.
We thank Dan Gohman for insightful discussions.
Conrad Watt is supported by an EPSRC Doctoral Training award, the
Semantic Foundations for Interactive Programs EPSRC program grant (\grantnum{EPSRC}{EP/N02706X/1}), and the
REMS: Rigorous Engineering for Mainstream Systems EPSRC program grant \grantnum{EPSRC}({EP/K008528/1}).
This work was supported in part by a gift from Cisco and by the CONIX Research
Center, one of six centers in JUMP, a Semiconductor Research Corporation
(SRC) program sponsored by DARPA.

We also thank Christopher Bell and the staff of NYCOS for their part in enabling this work to be completed on schedule.
\end{acks}

\bibliography{bibliography}

\end{document}